\documentclass[11pt]{article}

\usepackage[T1]{fontenc}
\usepackage{lmodern}
\usepackage[margin=1.08in]{geometry}
\usepackage{microtype}
\usepackage{amsmath,amssymb,amsthm,mathtools}
\usepackage{mathrsfs}
\usepackage{bm}
\usepackage{booktabs}
\usepackage{array}
\usepackage{float}
\usepackage{needspace}
\usepackage{enumitem}
\usepackage{xcolor}
\usepackage{tikz}
\usetikzlibrary{arrows.meta,positioning,calc,decorations.pathmorphing,fit,matrix}
\usepackage{tikz-cd}
\usepackage{hyperref}
\usepackage[capitalise,noabbrev]{cleveref}

\hypersetup{
  colorlinks=true,
  linkcolor=blue!45!black,
  citecolor=green!35!black,
  urlcolor=blue!55!black,
  pdftitle={Quantum error correcting codes from quiver moduli: Grassmannians, cyclic monodromy, and argyle collision divisors},
  pdfauthor={Steven Rayan and Evan Sundbo}
}

\setlist[itemize]{leftmargin=2em,itemsep=2pt,topsep=4pt}
\setlist[enumerate]{leftmargin=2.2em,itemsep=2pt,topsep=4pt}
\renewcommand{\arraystretch}{1.12}
\allowdisplaybreaks

\newtheorem{theorem}{Theorem}[section]
\newtheorem{proposition}[theorem]{Proposition}
\newtheorem{lemma}[theorem]{Lemma}
\newtheorem{corollary}[theorem]{Corollary}
\theoremstyle{definition}
\newtheorem{definition}[theorem]{Definition}
\newtheorem{example}[theorem]{Example}

\theoremstyle{remark}
\newtheorem{remark}[theorem]{Remark}

\newcommand{\C}{\mathbb C}

\newcommand{\Z}{\mathbb Z}
\newcommand{\Pp}{\mathbb P}
\newcommand{\cH}{\mathcal H}
\newcommand{\cK}{\mathcal K}

\newcommand{\cB}{\mathcal B}
\newcommand{\cA}{\mathcal A}

\newcommand{\cF}{\mathcal F}
\newcommand{\cR}{\mathcal R}
\newcommand{\cO}{\mathcal O}
\newcommand{\cC}{\mathcal C}
\newcommand{\cG}{\mathcal G}
\newcommand{\cL}{\mathcal L}
\newcommand{\End}{\operatorname{End}}
\newcommand{\Hom}{\operatorname{Hom}}

\newcommand{\Sym}{\operatorname{Sym}}
\newcommand{\Gr}{\operatorname{Gr}}

\newcommand{\Tr}{\operatorname{Tr}}

\newcommand{\Span}{\operatorname{span}}
\newcommand{\Ind}{\operatorname{Ind}}
\newcommand{\Res}{\operatorname{Res}}
\newcommand{\diag}{\operatorname{diag}}

\newcommand{\reg}{\mathrm{reg}}
\newcommand{\std}{\mathrm{std}}
\newcommand{\sgn}{\mathrm{sgn}}
\newcommand{\ket}[1]{\lvert #1\rangle}
\newcommand{\bra}[1]{\langle #1\rvert}

\newcommand{\proj}[1]{\lvert #1\rangle\!\langle #1\rvert}

\title{\textbf{Quantum error correcting codes from quiver moduli: Grassmannians, cyclic monodromy, and argyle collision divisors}}
\author{
Steven Rayan\thanks{Centre for Quantum Topology and Its Applications (quanTA) and Department of Mathematics and Statistics, University of Saskatchewan, Saskatoon, Saskatchewan S7N 5E6, Canada. \texttt{rayan@math.usask.ca}}
\and
Evan Sundbo\thanks{Department of Pure Mathematics, University of Waterloo, Canada. \texttt{e2sundbo@uwaterloo.ca}}
}
\date{November 3, 2026}

\begin{document}
\maketitle

\begin{abstract}
Building on aspects of a recent diagrammatic field-theory programme for quantum error correction put forward by the first-named author in \cite{RayanDiagrammatic2026}, we construct exact finite-dimensional codes from twisted cyclic and argyle quiver moduli.  The regular fibre of the node-labelled cyclic Hitchin map on the projective line is the set of allocations of $M=rt$ simple spectral zeroes among arrows of degrees $\boldsymbol m=(m_1,\ldots,m_r)$.  Its sheet Hilbert space is the Young permutation module
\[
 \cH_{\boldsymbol m}\cong
 \Ind_{S_{m_1}\times\cdots\times S_{m_r}}^{S_M}\mathbf1.
\]
Root monodromy generates the full matrix algebra on each Specht factor and acts trivially on its Kostka multiplicity factor.  Gauge reset gives a channel-independent exact recovery.  We prove that the largest dimension of any subspace code correcting the full monodromy algebra, including codes spanning several isotypic blocks, is $\max_\lambda K_{\lambda\boldsymbol m}$.

An explicitly constructed argyle master space quantizes to $(\C^r)^{\otimes M}$; its weight-$\boldsymbol m$ sector identifies equivariantly with the cyclic sheet space.  This identification descends through the ordered-root cover to an isomorphism of associated unitary local systems.  For the balanced three-cycle, we compute the complete spectrum
\[
 K_{(a,b,c),(t,t,t)}=1+\min(a-b,b-c),
 \qquad a+b+c=3t,
\]
and deduce the optimal exact logical dimension $t+1$, attained uniquely among isotypic blocks by $(2t,t)$.  At $t=1$ this gives explicit codewords for a six-sheet logical qubit.  The algebra generated by its commuting right action realizes every logical unitary, exhibiting errors with identical monodromy-footprint statistics but incompatible exact recoveries.  We also calculate collision-parity ranks, distinguish safe quantum footprints from classical collision tests, give a higher-rank $\Gr(2,6)$ twisted-chain example, and prove both a conditional criterion for stratified collision recovery and an operational perturbation bound.
\end{abstract}



\setcounter{tocdepth}{2}
\tableofcontents

\section{Introduction}

Quantum error correction is an exercise in controlled information loss.  An error history contains more microscopic information than a recovery operation should attempt to learn.  The measured syndrome is deliberately coarse: it must distinguish enough error sectors to permit recovery, but it must not reveal the encoded state.  In a surface code, an error chain is observed through its boundary and remains ambiguous up to a cycle.  In a field-theoretic code, a defect history is observed through a local charge, fusion channel, or boundary datum.  The term \emph{footprint} was introduced in \cite{RayanDiagrammatic2026} for this intermediate datum, with a measured, logically safe footprint playing the role of a syndrome.

This article builds on specific aspects of the programme put forward there by the first-named author.  In particular, we use the distinction between an error history, its measured footprint, and the residual ambiguity inside a footprint fibre, together with the fibrewise Knill--Laflamme criterion, as design principles for a quiver-moduli code.  The construction addresses the regular-fibre part of the Higgs-bundle and spectral-curve direction posed in \cite[Problem~11.12]{RayanDiagrammatic2026}, providing a Hilbert space, encoding, error algebra, logically safe measurements, and recovery.  It does not require a quantization of singular spectral curves or their compactified Jacobians.  For a complementary exposition connecting operational, diagrammatic, error-correcting, and geometric viewpoints on quantum circuits, see the first-named author's lecture notes \cite{RayanRethinking2026}.

The purpose of this article is to show that twisted quiver moduli on $\Pp^1$ supply a finite and computable realization of this pattern.  The relevant geometric structures already appear in two earlier works of the present authors.  Twisted argyle quivers supply projective-space and Grassmannian descriptions of moduli strata and their completions; their boundaries are governed by collisions of gauge-invariant polynomial zeroes and by changes in the Birkhoff--Grothendieck splitting type \cite{RayanSundboArgyle2018}.  The node-labelled moduli of twisted cyclic quivers have finite regular Hitchin fibres obtained by distributing the zeroes of a spectral polynomial among the arrows; the covering ramifies when zeroes collide \cite{RayanSundboCyclic2021}.  To define a code from this covering, we specify a Hilbert space, an encoding, an error algebra, a measurement model, and a recovery operation.

Let $Q$ be a cyclic quiver with $r$ line-bundle nodes, let the twist be $\cO_{\Pp^1}(t)$, and write
\[
  m_i=d_{i+1}-d_i+t,\qquad i\in\Z/r\Z.
\]
The integers $m_i$ are the degrees of the arrow polynomials and satisfy $\sum_i m_i=rt$.  Over a Hitchin-base point $\gamma$ with $rt$ distinct zeroes, a sheet of the cyclic covering is an ordered partition of the root set into blocks of sizes $m_1,\ldots,m_r$.  Passing from the finite sheet set to its $\ell^2$-space gives
\[
 \cH_{\boldsymbol m}
 \cong \C\big[S_{rt}/(S_{m_1}\times\cdots\times S_{m_r})\big].
\]
Root monodromy acts through $S_{rt}$.  Thus the geometric error algebra is the image of $\C[S_{rt}]$.

Young's rule decomposes this permutation module as
\begin{equation}\label{eq:intro-young}
 \cH_{\boldsymbol m}
 \cong
 \bigoplus_{\lambda\vdash rt} S^\lambda\otimes \C^{K_{\lambda\boldsymbol m}},
\end{equation}
where $S^\lambda$ is a Specht module and $K_{\lambda\boldsymbol m}$ is a Kostka number.  Monodromy acts on the first tensor factor and trivially on the second.  The multiplicity factor is therefore a noiseless subsystem in the standard operator-QEC sense \cite{KnillLaflamme1997,KribsLaflammePoulinLesosky2006,NielsenPoulin2007}.  More strongly, after fixing a unit vector in $S^\lambda$, the resulting copy of $\C^{K_{\lambda\boldsymbol m}}$ is an ordinary subspace code satisfying the Knill--Laflamme equations for the entire monodromy algebra.  Recovery simply discards and resets the Specht gauge factor.

The same decomposition arises from quantizing an argyle quiver.  For $M=rt$, we construct a stable rank-one argyle chain whose moduli space is
\[
  X_{M,r}\cong (\Pp^{r-1})^M.
\]
Quantization by $\cO(1,\ldots,1)$ gives
\[
  \cK_{M,r}=H^0(X_{M,r},\cO(1,\ldots,1))^*
  \cong (\C^r)^{\otimes M}.
\]
The torus-weight-$\boldsymbol m$ sector has the word basis with content $\boldsymbol m$, and this basis is precisely the set of cyclic root allocations.  Consequently
\begin{equation}\label{eq:intro-bridge}
  \cK_{M,r}[\boldsymbol m]\cong \cH_{\boldsymbol m}
\end{equation}
equivariantly for $S_M$.  Schur--Weyl duality on the argyle quantization restricts to Young's rule on the cyclic fibre.  The protected Kostka spaces are the weight spaces of the corresponding polynomial $\mathrm{GL}_r$-modules.

This bridge produces a small exact code and a growing family.  For the balanced three-cycle with $t=1$, the generic fibre has six sheets and is the regular $S_3$-module.  Its standard isotypic component is
\[
  S^{(2,1)}\otimes\C^2,
\]
so fixing a gauge vector yields one protected logical qubit.  For arbitrary $t\geq1$, the balanced composition $(t,t,t)$ and partition $(2t,t)$ satisfy
\[
  K_{(2t,t),(t,t,t)}=t+1.
\]
In fact, for every $\lambda=(a,b,c)\vdash3t$ with at most three rows,
\[
 K_{\lambda,(t,t,t)}=1+\min(a-b,b-c)\leq t+1,
\]
with equality only for $\lambda=(2t,t,0)$; all longer partitions have zero multiplicity.  Combined with a general upper bound for subspace codes of a finite-dimensional error algebra, this proves that $t+1$ is the optimal exact logical dimension in the entire balanced sheet space, even if an encoding is allowed to span several isotypic blocks.  The familiar weight-multiplicity calculation thus yields a sharp coding theorem for this geometric error model.  The model does not claim protection from arbitrary local qutrit noise.  Its error algebra is geometrically selected and correlated: it consists of monodromy and collision-permutation operators.

Collision geometry also supplies measurements.  A simple collision exchanges two roots and gives a transposition $\tau$.  The projectors $(I\pm\rho(\tau))/2$ measure collision parity.  Projectors for disjoint collisions commute, whereas those for overlapping collisions need not.  The Young--Jucys--Murphy elements give a canonical commuting family whose joint eigenvectors are standard tableaux.  On the code block these operators act only on the gauge factor, so their spectral projectors reveal no logical information.  They are syndrome-admissible whenever the selected error representatives are resolved by their joint spectrum.

The commutant of the monodromy algebra acts on the Kostka multiplicity factor.  Two errors that differ by such a commutant unitary have identical statistics for every monodromy-footprint measurement, yet their residual difference may act logically.  In the six-sheet example, left multiplication by $S_3$ is the root-monodromy algebra and right multiplication is its commuting colour-relabeling algebra.  We exhibit two same-footprint errors differing by a logical Pauli operation.  This is the quiver-moduli counterpart of a nontrivial footprint fibre in topological decoding.

The decomposition in \eqref{eq:intro-young} is Young's rule \cite{FultonYoungTableaux1997,Sagan2001}.  Noiseless subsystems \cite{KribsLaflammePoulinLesosky2006} and the quantum Schur transform \cite{BaconChuangHarrow2006} are also standard.  The contribution here is their geometric assembly and its exact consequences: cyclic Hitchin fibres become physical sheet spaces, argyle moduli provide a quantized master space, discriminant monodromy becomes the error algebra, collision operators become footprint observables, and Kostka multiplicities become logical systems.  The algebra theorem identifies the complete Wedderburn pair consisting of the monodromy algebra and its commutant.  The resulting optimal-dimension theorem bounds all exact subspace encodings, not only gauge-fixed ones, and the complete balanced spectrum makes that bound explicit.  The collision-character calculation determines the size of every binary collision sector, while the balanced Kostka theorem produces an explicit hierarchy of codes rather than a single isolated example.  For the six-sheet qubit, the compressed right $S_3$-action realizes the full logical algebra while leaving the monodromy-footprint statistics unchanged.

Related descriptions of Higgs fields in terms of divisors appear in the work of Dinh and Teschner \cite{DinhTeschner2023}.  For a stable rank-two Higgs bundle on a curve of genus at least two, they write the Higgs field in a local frame adapted to a line subbundle.  When the spectral curve is smooth, the associated Baker--Akhiezer divisor projects to the zero divisor of the lower-left entry.  More recently, Dinh, Figiel, and Teschner \cite{DinhFigielTeschner2026} study changes of stratum through collisions of apparent singularities of opers and describe analogous Higgs-bundle families using Baker--Akhiezer divisors.  This provides a higher-genus comparison for the role of component zeroes and collision strata in the quiver models considered here.

In a related setting over $\Pp^1$, Dinh and the first-named author \cite{DinhRayan2026} identify Baker--Akhiezer sections of rank-two $\cO_{\Pp^1}(3)$-twisted Higgs bundles with states in the lowest Landau level on genus-two spectral curves in a uniform magnetic field of two flux quanta.  The limiting vortex divisors describe the change of splitting type at the Hitchin section and the associated jump in the dimension of the lowest Landau level.

\subsection*{Organization and main results}

Section~\ref{sec:prelim} recalls subsystem QEC, cyclic and argyle quivers, and the operational footprint distinction.  Section~\ref{sec:cyclic-fibres} identifies regular cyclic Hitchin fibres with root-allocation permutation modules, computes their monodromy, determines the monodromy algebra and commutant, proves the exact code theorem, and computes collision and Young--Jucys--Murphy footprint sectors.  Section~\ref{sec:argyle-master} builds the argyle master space and globalizes the equivariant weight-sector bridge \eqref{eq:intro-bridge}.  Section~\ref{sec:six-sheet} gives the six-sheet one-qubit code, its complete logical commutant and same-footprint ambiguity, and the complete spectrum and optimal logical dimension of the balanced three-cycle family.  Section~\ref{sec:higher-argyle} gives a higher-rank Grassmannian example and a stratified collision criterion, makes the encoding and recovery explicit, gives a perturbative test for errors outside the exact algebra, and records the resulting field-theoretic footprint dictionary.

\section{Preliminaries}\label{sec:prelim}

\subsection{Subspace and subsystem quantum error correction}

Let $\cH_P$ be a finite-dimensional physical Hilbert space and let $\cC\subset\cH_P$ be a code subspace with orthogonal projector $P$.  A finite error family $\{E_a\}$ is exactly correctable if and only if
\begin{equation}\label{eq:KL}
  PE_a^\dagger E_bP=\lambda_{ab}P
\end{equation}
for a positive semidefinite matrix $(\lambda_{ab})$ \cite{KnillLaflamme1997}.  By linearity, the condition applies to the span of the specified errors.

Subsystem QEC begins with a decomposition
\begin{equation}\label{eq:subsystem-decomposition}
 \cH_P=(\cH_G\otimes\cH_L)\oplus\cH_R,
\end{equation}
where $\cH_L$ carries the logical information and $\cH_G$ is a gauge factor.  If every error in an algebra $\cA$ restricts to
\begin{equation}\label{eq:gauge-algebra}
  A\big|_{\cH_G\otimes\cH_L}=A_G\otimes I_L,
\end{equation}
then $\cH_L$ is noiseless for every channel with Kraus operators in $\cA$ \cite{KribsLaflammePoulinLesosky2006}.  Fixing a unit gauge vector $g_0\in\cH_G$ produces the ordinary code subspace
\begin{equation}\label{eq:gauge-fixed-code}
  \cC_{g_0}=\C g_0\otimes\cH_L.
\end{equation}
It satisfies \eqref{eq:KL} for the whole algebra $\cA$, because
\begin{equation}\label{eq:gauge-fixed-KL}
 P_{g_0}(A_G^\dagger B_G\otimes I_L)P_{g_0}
 =\bra{g_0}A_G^\dagger B_G\ket{g_0}\,P_{g_0}.
\end{equation}
This elementary observation will be the QEC mechanism used throughout.

\subsection{Twisted cyclic and argyle quivers on \texorpdfstring{$\Pp^1$}{the projective line}}

Fix $L=\cO_{\Pp^1}(t)$ with $t>0$.  A type $(1,\ldots,1)$ twisted cyclic quiver with $r$ nodes is a diagram
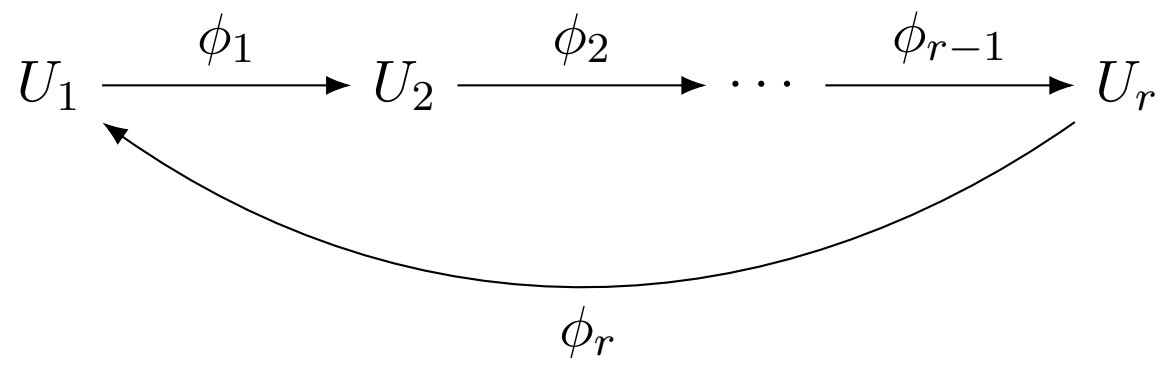
\begin{equation}\label{eq:cyclic-quiver}
\begin{tikzpicture}[baseline=-2pt,>=Latex,node distance=1.65cm]
\node (u1) {$U_1$};
\node (u2) [right=of u1] {$U_2$};
\node (ud) [right=of u2] {$\cdots$};
\node (ur) [right=of ud] {$U_r$};
\draw[->] (u1) -- node[above] {$\phi_1$} (u2);
\draw[->] (u2) -- node[above] {$\phi_2$} (ud);
\draw[->] (ud) -- node[above] {$\phi_{r-1}$} (ur);
\draw[->] (ur) to[bend left=35] node[below] {$\phi_r$} (u1);
\end{tikzpicture}
\end{equation}
with $U_i=\cO_{\Pp^1}(d_i)$ and
\[
\phi_i\in H^0\big(\Pp^1,U_i^*\otimes U_{i+1}\otimes L\big),
 \qquad U_{r+1}=U_1.
\]
Set
\begin{equation}\label{eq:mi-def}
 m_i=d_{i+1}-d_i+t.
\end{equation}
Then $\phi_i$ is represented by a homogeneous polynomial of degree $m_i$, and
\begin{equation}\label{eq:sum-mi}
  \sum_{i=1}^r m_i=rt=:M.
\end{equation}
We assume $r\geq2$ and $m_i\geq0$.  Throughout, cyclic moduli retain the labelled quiver grading: an isomorphism is a tuple of bundle isomorphisms $U_i\to U_i'$ intertwining the arrows.  We do not quotient by permutations of the nodes or by arbitrary automorphisms of the ungraded bundle $\bigoplus_iU_i$.  In particular, the sheet count below is a count in graded quiver moduli.  Forgetting the grading need not be injective into the ambient Higgs moduli space.  Over a nonzero Hitchin-base point all arrows are nonzero; over a reduced zero divisor, stability follows directly from \Cref{lem:regular-cyclic-stability} below.

The associated Higgs field on $E=\bigoplus_iU_i$ is the cyclic matrix with the $\phi_i$ on the cyclic superdiagonal.  This is the cyclic restriction of the Hitchin map and its spectral construction \cite{Hitchin1987,BeauvilleNarasimhanRamanan1989}.  Up to a convention-dependent sign, its only nonzero characteristic coefficient is
\begin{equation}\label{eq:cyclic-hitchin}
  h(E,\Phi)=\phi_1\phi_2\cdots\phi_r
  \in H^0(\Pp^1,\cO(M)).
\end{equation}
Thus the graded cyclic Hitchin map takes values in the $H^0(\Pp^1,\cO(M))$ summand of the Hitchin base.  The regular-fibre factorization in the proof of Theorem~5.1 (arXiv version of \cite{RayanSundboCyclic2021}) gives the covering degree
\begin{equation}\label{eq:eta}
  \eta(\boldsymbol m)=\binom{M}{m_1,\ldots,m_r}
  =\frac{M!}{m_1!\cdots m_r!}.
\end{equation}

\paragraph{Argyle quivers and collision strata.}

An argyle quiver is an $A$-type quiver whose ranks alternate between $1$ and arbitrary positive integers:
\begin{equation}\label{eq:argyle-shape}
  1\longrightarrow k_1\longrightarrow 1\longrightarrow k_2
  \longrightarrow\cdots\longrightarrow k_s\longrightarrow1.
\end{equation}
For fixed Birkhoff--Grothendieck splitting types at the higher-rank nodes, \cite{RayanSundboArgyle2018} describes collision-free strata and their projective completions using projective spaces and Grassmannians.  The block decomposition for a general argyle quiver uses stability induced from the entire chain, not necessarily the ordinary stability of each isolated $(1,k_i,1)$ block.  In Section~\ref{sec:higher-argyle}, tensor-product quantization is stated with its global product and polarization hypotheses explicit.

The nonreductive part of the bundle-automorphism group acts on polynomial entries by Euclidean reduction.  Some polynomial zeroes survive this reduction and are invariants of the isomorphism class.  When prescribed incoming and outgoing invariant zeroes collide, a meromorphic change of frame can alter the splitting type by an elementary move
\begin{equation}\label{eq:type-change}
 (a_1,\ldots,a_i,\ldots,a_j,\ldots,a_k)
 \longmapsto
 (a_1,\ldots,a_i+1,\ldots,a_j-1,\ldots,a_k).
\end{equation}
The stable splitting types and the allowed moves \eqref{eq:type-change} form a finite type-change graph.  We return to its QEC role in Section~\ref{sec:higher-argyle}.

\subsection{Footprints and syndrome admissibility}

We use the following finite-dimensional specialization of the footprint language of \cite{RayanDiagrammatic2026}.

\begin{definition}\label{def:footprint-algebra}
Let $\cC\subset\cH_P$ be a code with projector $P$, and let
\[
 \cF=\Span\{\Pi_s:s\in S\}
\]
be a commutative algebra generated by orthogonal projectors summing to the identity.
It is a \emph{diagnostic footprint algebra} for an error family $\{E_a\}$ if its outcomes record a chosen coarse property of the errors.  It is \emph{no-error safe} if there is an $s_0$ such that
\[
 \Pi_{s_0}P=P,\qquad \Pi_sP=0\quad(s\neq s_0).
\]
It is \emph{syndrome-admissible} for chosen representatives if it is no-error safe, its measurement reveals no logical information, and every representative maps the code into a specified measured sector.  The \emph{footprint fibre} over $s$ is the set of representatives assigned to that sector.
\end{definition}

Different representatives in one footprint fibre are harmless precisely when their residual composites satisfy the Knill--Laflamme scalar condition.  In the present article, collision projectors act as $\Pi_s^G\otimes I_L$ on the selected code block.  They therefore reveal gauge information only for an independently prepared gauge state, as in the gauge-fixed encoding.  Logical non-leakage and no-error safety are distinct requirements: a gauge-only measurement is no-error safe precisely when the prepared gauge vector lies in one of its outcome sectors.

\section{Cyclic fibres and exact monodromy codes}\label{sec:cyclic-fibres}\label{sec:codes}

\subsection{Root allocations and braid monodromy}

Let
\begin{equation}\label{eq:regular-base}
 \cB_M^{\reg}
 =\{\gamma\in H^0(\Pp^1,\cO(M)):\gamma\neq0,
       \text{ and }\operatorname{div}(\gamma)\text{ is reduced}\}.
\end{equation}
For $\gamma\in\cB_M^{\reg}$, write its zero set as an unordered set
\[
 Z(\gamma)=\{z_1,\ldots,z_M\}\subset\Pp^1.
\]
An ordering will be chosen only to identify a fibre with a standard permutation module; changing the ordering conjugates all resulting actions.

\begin{lemma}[Stability over a reduced spectral divisor]\label{lem:regular-cyclic-stability}
If $\gamma\in\cB_M^{\reg}$ and $\phi_1\cdots\phi_r=\gamma$, the associated twisted Higgs bundle is stable.  No coprimality condition on $r$ and $\sum_i d_i$ is needed.
\end{lemma}

\begin{proof}
Choose a simple zero $z$ of $\gamma$ and trivialize the twisting line bundle near $z$.  Over the local discrete valuation ring $\cO_{\Pp^1,z}$, the characteristic polynomial of the cyclic Higgs matrix is $x^r-\gamma$, up to the fixed sign convention.  Its constant coefficient has valuation one, so Eisenstein's criterion makes it irreducible over $\C(\Pp^1)$.  After trivializing the twist over this function field, the generic Higgs endomorphism therefore has no nonzero proper invariant subspace: its underlying module is one-dimensional over the degree-$r$ field extension defined by $x^r-\gamma$.  Any proper invariant subbundle would give such a subspace.  Hence the stability inequalities are vacuous.
\end{proof}

The cyclic moduli discussion in \cite{RayanSundboCyclic2021} imposes a coprimality hypothesis in its general setup.  The preceding argument and the direct factorization proof below establish the regular-fibre assertions used here also in the noncoprime balanced cases.

\begin{definition}\label{def:allocation-set}
For a composition $\boldsymbol m=(m_1,\ldots,m_r)$ of $M$, the \emph{root-allocation set} is
\begin{equation}\label{eq:allocation-set}
 \Omega_{\boldsymbol m}(Z)
 =\left\{(B_1,\ldots,B_r):
 Z=B_1\sqcup\cdots\sqcup B_r,\ |B_i|=m_i\right\}.
\end{equation}
Its elements will also be called coloured root allocations, with the elements of $B_i$ carrying colour $i$.
\end{definition}

\begin{proposition}[Cyclic fibre-allocation correspondence]\label{prop:fibre-allocation}
Let $Q$ be a stable type $(1,\ldots,1)$ cyclic quiver on $\Pp^1$ with arrow degrees $\boldsymbol m$, and let $\gamma\in\cB_M^{\reg}$.  There is a natural bijection
\begin{equation}\label{eq:fibre-allocation-bijection}
 h_Q^{-1}(\gamma)\cong\Omega_{\boldsymbol m}(Z(\gamma)).
\end{equation}
In particular, the fibre contains $\eta(\boldsymbol m)$ points.
\end{proposition}

\begin{proof}
For a cyclic representation over $\gamma$, equation \eqref{eq:cyclic-hitchin} gives
\[
  \phi_1\cdots\phi_r=\gamma
\]
after absorbing the fixed sign into one arrow.  Since $\gamma$ has simple zeroes, every zero occurs in exactly one factor.  Setting
\[
  B_i=\operatorname{div}(\phi_i)_{\mathrm{red}}
\]
therefore gives an ordered partition of $Z(\gamma)$ with $|B_i|=m_i$.

Conversely, an allocation $(B_1,\ldots,B_r)$ determines each $\phi_i$ up to a nonzero scalar by its degree-$m_i$ divisor.  The product of those scalars is fixed by the equality with $\gamma$.  The effective automorphism group $(\C^*)^{r-1}$ of the line-bundle representation acts transitively on scalar tuples with the prescribed product.  Thus each allocation defines exactly one node-labelled moduli point, and \Cref{lem:regular-cyclic-stability} guarantees its stability.  The count is the multinomial coefficient \eqref{eq:eta}.
\end{proof}

\begin{definition}\label{def:sheet-Hilbert}
The \emph{cyclic sheet Hilbert space} over $\gamma$ is
\begin{equation}\label{eq:sheet-Hilbert}
 \cH_{\gamma,\boldsymbol m}=\ell^2\big(h_Q^{-1}(\gamma)\big),
\end{equation}
with the moduli sheets as an orthonormal basis.  Via \Cref{prop:fibre-allocation}, we also write it as $\ell^2(\Omega_{\boldsymbol m})$.
\end{definition}

The passage from a finite fibre to its $\ell^2$-space is a finite quantization choice.  Unlike a cohomological or deformation quantization of the whole moduli space, it introduces no semiclassical parameter: distinct sheets are simply declared orthogonal.  This is the appropriate Hilbertization for the finite-cover error model studied below.

\begin{proposition}[Young permutation module]\label{prop:young-permutation-module}
Choose an ordering of $Z(\gamma)$ and let
\[
 S_{\boldsymbol m}=S_{m_1}\times\cdots\times S_{m_r}\subseteq S_M
\]
be the Young subgroup stabilizing the standard allocation.  Then
\begin{equation}\label{eq:young-permutation-module}
 \cH_{\gamma,\boldsymbol m}
 \cong \C[S_M/S_{\boldsymbol m}]
 \cong \Ind_{S_{\boldsymbol m}}^{S_M}\mathbf1
\end{equation}
as unitary $S_M$-modules.
\end{proposition}

\begin{proof}
The symmetric group acts transitively on ordered allocations by permuting the roots.  The stabilizer of the standard allocation consists of the permutations internal to each block, namely $S_{\boldsymbol m}$.  Linearizing the resulting $S_M$-set gives \eqref{eq:young-permutation-module}.
\end{proof}

\paragraph{Braid monodromy and the discriminant.}

Projectivizing $\cB_M^{\reg}$ removes the irrelevant nonzero scalar multiplying $\gamma$.  The zero-divisor map identifies the result with the unordered configuration space of $M$ points on $\Pp^1$.  Its fundamental group is the spherical braid group.  On an affine chart avoiding $\infty$, one obtains the usual Artin braid group $B_M$.  In either description, the permutation of strands gives a surjection to $S_M$.

\begin{theorem}[Monodromy of the cyclic covering]\label{thm:monodromy}
The graded cyclic covering descends along scalar rescaling to $\Pp\cB_M^{\reg}$, where its monodromy is
\begin{equation}\label{eq:monodromy-factorization}
 \pi_1(\Pp\cB_M^{\reg},[\gamma])
 \longrightarrow S_M
 \xrightarrow{\ \rho_{\boldsymbol m}\ }
 U(\cH_{\gamma,\boldsymbol m}),
\end{equation}
where $\rho_{\boldsymbol m}$ is the permutation representation on $S_M/S_{\boldsymbol m}$.  The monodromy group is the full image $\rho_{\boldsymbol m}(S_M)$.
\end{theorem}

\begin{proof}
The allocation cover depends only on the zero divisor, so over $\cB_M^{\reg}$ it is pulled back from the projectivized base; loops in the scalar $\C^*$ fibre act trivially.  A loop in the projectivized regular base moves the simple zeroes of $\gamma$ without collision and returns their unordered set to itself.  Continuing the factors $\phi_i$ along the loop transports the colour attached to each zero.  If the induced root permutation is $\sigma$, the terminal allocation is
\[
 (B_1,\ldots,B_r)\longmapsto
 (\sigma B_1,\ldots,\sigma B_r),
\]
which is exactly $\rho_{\boldsymbol m}(\sigma)$.  Simple half-twists of adjacent roots map to adjacent transpositions, which generate $S_M$.  Hence the image is the full permutation image.
\end{proof}

The theorem makes the error histories geometric.  A path in the discriminant complement is a braid history; the corresponding ideal error is a unitary sheet permutation.  The representation factors through $S_M$, so pure braids act trivially in this finite sheet model.  The exact code studied here is therefore determined by permutation monodromy alone.

For the boundary calculation, use the finite multiplication map
\begin{equation}\label{eq:allocation-completion}
 \prod_{i=1}^r\Pp H^0(\Pp^1,\cO(m_i))
 \longrightarrow\Pp H^0(\Pp^1,\cO(M)),
 \qquad ([\phi_i])_i\longmapsto[\phi_1\cdots\phi_r].
\end{equation}
It is proper with finite fibres, since a fixed divisor has only finitely many decompositions of the prescribed degrees.  Its source is smooth and its restriction over the regular base is the allocation cover.  Thus it supplies the normal finite extension used below.  Boundary points of this extension are not asserted to be stable Higgs bundles; in noncoprime cases some can be strictly semistable.

Let $\Delta\subset\Pp H^0(\Pp^1,\cO(M))$ be the discriminant hypersurface.  Near a smooth point of $\Delta$, exactly two zeroes coalesce.  A small loop around $\Delta$ exchanges them and gives a transposition $\tau=(ab)$.

\begin{proposition}[Local collision structure]\label{prop:local-collision}
Let $\tau=(ab)$ be the transposition associated with a simple collision of roots $z_a,z_b$.  On $\Omega_{\boldsymbol m}$:
\begin{enumerate}
\item an allocation is fixed by $\tau$ exactly when $z_a$ and $z_b$ have the same colour;
\item allocations in which the two roots have different colours occur in two-element $\tau$-orbits;
\item in the finite extension \eqref{eq:allocation-completion}, the corresponding paired sheets have local ramification index $2$ along the collision divisor.
\end{enumerate}
\end{proposition}

\begin{proof}
The first two statements are immediate from the action on coloured allocations.  For different colours, interchanging the two roots produces a distinct allocation and applying the interchange twice returns to the original one.  At a smooth discriminant point, a transverse coordinate can be chosen so that the unordered root parameter is $u^2$, where $u$ changes sign under the half-twist.  The two exchanged allocation branches are therefore locally joined by $u\mapsto u^2$, giving ramification index $2$.  Same-colour allocations are fixed and do not form such a paired branch.
\end{proof}

The same/different-colour datum describes the classical ramification pattern.  Its diagonal measurement on the sheet basis need not preserve a chosen isotypic block or act only on its gauge factor.  The quantum observables used below are instead the spectral projectors of collision monodromy; their logical safety is proved separately.

\subsection{Young decomposition and protected multiplicities}

Let $\mu$ be the partition obtained by arranging the nonzero parts of $\boldsymbol m$ in weakly decreasing order.  Reordering the colours conjugates the Young subgroup and does not change the representation.  For a partition $\lambda\vdash M$, let $S^\lambda$ be the corresponding complex Specht module, let $f^\lambda=\dim S^\lambda$, and let $K_{\lambda\mu}$ be the Kostka number.  We also write $K_{\lambda\boldsymbol m}$ because Kostka multiplicity is insensitive to a permutation of the content.

Young's rule gives a unitary isotypic decomposition \cite{FultonYoungTableaux1997,Sagan2001}
\begin{equation}\label{eq:young-rule}
 \cH_{\gamma,\boldsymbol m}
 \cong
 \bigoplus_{\substack{\lambda\vdash M\\ K_{\lambda\boldsymbol m}>0}}
 S^\lambda\otimes \cL_{\lambda,\boldsymbol m},
 \qquad
 \dim\cL_{\lambda,\boldsymbol m}=K_{\lambda\boldsymbol m}.
\end{equation}
Here $S_M$ acts on $S^\lambda$ and trivially on the multiplicity space $\cL_{\lambda,\boldsymbol m}$.
The summand occurs precisely when $\lambda$ dominates the partition $\mu$ obtained from $\boldsymbol m$.  Multiplicity one therefore gives a gauge block but no nontrivial logical system, whereas $K_{\lambda\boldsymbol m}>1$ is the exact combinatorial condition for a protected logical degree of freedom in that block.

\begin{definition}\label{def:monodromy-error-algebra}
The \emph{cyclic monodromy error algebra} is
\begin{equation}\label{eq:monodromy-error-algebra}
 \cA_{\mathrm{mon}}(\boldsymbol m)
 =\rho_{\boldsymbol m}(\C[S_M])
 \subseteq\End(\cH_{\gamma,\boldsymbol m}).
\end{equation}
A \emph{monodromy channel} is a completely positive trace-preserving map whose Kraus operators belong to $\cA_{\mathrm{mon}}(\boldsymbol m)$.
\end{definition}

This includes stochastic mixtures of braid monodromies, coherent linear combinations of sheet permutations, and channels obtained by measuring collision-parity operators.  It does not include arbitrary operators on the sheet Hilbert space.

\begin{theorem}[Monodromy algebra and logical commutant]\label{thm:algebra-commutant}
Let
\[
 \Lambda_{\boldsymbol m}
 =\{\lambda\vdash M:K_{\lambda\boldsymbol m}>0\}.
\]
Relative to \eqref{eq:young-rule}, the monodromy algebra and its commutant are
\begin{align}
 \cA_{\mathrm{mon}}(\boldsymbol m)
 &=\bigoplus_{\lambda\in\Lambda_{\boldsymbol m}}
   \End(S^\lambda)\otimes I_{\cL_{\lambda,\boldsymbol m}},
   \label{eq:mon-algebra-wedderburn}\\
 \cA_{\mathrm{mon}}(\boldsymbol m)'
 &=\bigoplus_{\lambda\in\Lambda_{\boldsymbol m}}
   I_{S^\lambda}\otimes\End(\cL_{\lambda,\boldsymbol m}).
   \label{eq:mon-commutant-wedderburn}
\end{align}
In particular, every linear logical operator on a Kostka factor is the restriction of a global operator commuting with monodromy.  The orthogonal projector onto the $\lambda$-isotypic block belongs to the monodromy algebra and is explicitly
\begin{equation}\label{eq:central-isotypic-projector}
 P_\lambda
 =\frac{f^\lambda}{M!}
   \sum_{\sigma\in S_M}\chi^\lambda(\sigma^{-1})
   \rho_{\boldsymbol m}(\sigma).
\end{equation}
\end{theorem}

\begin{proof}
Maschke's theorem and Young's rule give the isotypic decomposition \eqref{eq:young-rule}.  The image of $\C[S_M]$ on the irreducible module $S^\lambda$ is all of $\End(S^\lambda)$, either by Burnside's theorem or directly from the Artin--Wedderburn decomposition of the group algebra.  Equivalent copies are collected in the multiplicity factor, so the same matrix acts on every copy; this gives \eqref{eq:mon-algebra-wedderburn}.  Schur's lemma then gives \eqref{eq:mon-commutant-wedderburn}.  Formula \eqref{eq:central-isotypic-projector} is the image of the central primitive idempotent associated with $\lambda$.
\end{proof}

\begin{corollary}[Maximal protected factor]\label{cor:maximal-protected-factor}
Within the $\lambda$-isotypic block, $\cL_{\lambda,\boldsymbol m}$ is the maximal noiseless logical factor for the full monodromy algebra: every noiseless subsystem in that block has logical dimension at most $K_{\lambda\boldsymbol m}$.  Moreover, the full logical matrix algebra is exactly the corresponding block of $\cA_{\mathrm{mon}}(\boldsymbol m)'$.
\end{corollary}

\begin{proof}
The defining representation of the full matrix algebra $\End(S^\lambda)$ occurs with multiplicity $K_{\lambda\boldsymbol m}$.  Any subsystem on which this algebra acts only on a gauge tensor factor can use no more than that multiplicity.  Equality is attained by \eqref{eq:young-rule}, and the final assertion is \eqref{eq:mon-commutant-wedderburn}.
\end{proof}

\begin{theorem}[Optimal dimension for the full monodromy algebra]\label{thm:optimal-dimension}
The largest dimension of a nonzero subspace code satisfying the Knill--Laflamme equations for the entire algebra $\cA_{\mathrm{mon}}(\boldsymbol m)$ is
\begin{equation}\label{eq:optimal-dimension}
 k_{\max}(\boldsymbol m)=\max_{\lambda\in\Lambda_{\boldsymbol m}}K_{\lambda\boldsymbol m}.
\end{equation}
The bound also applies to the logical dimension of every exactly correctable subsystem for this full error algebra.
\end{theorem}

\begin{proof}
Let $V:\C^k\to\cH_{\gamma,\boldsymbol m}$ be an isometric encoder for such a subspace code.  Since the error algebra is unital, Knill--Laflamme implies $V^\dagger A V=\alpha(A)I_k$ for every $A\in\cA_{\mathrm{mon}}$.  Choose an orthonormal basis $\{e_{\lambda i}\}_i$ of each Specht factor and expand the encoder in components
\[
 V_{\lambda i}=(\bra{e_{\lambda i}}\otimes I)W_\lambda^\dagger V:
 \C^k\longrightarrow\cL_{\lambda,\boldsymbol m},
\]
where $W_\lambda$ includes the isotypic block.  The diagonal matrix units on the Specht factors belong to the algebra by \eqref{eq:mon-algebra-wedderburn}, so
\[
 V_{\lambda i}^\dagger V_{\lambda i}=c_{\lambda i}I_k,
 \qquad c_{\lambda i}\geq0,
 \qquad\sum_{\lambda,i}c_{\lambda i}=1.
\]
For some $(\lambda,i)$ the scalar is positive, and $c_{\lambda i}^{-1/2}V_{\lambda i}$ is an isometric embedding into $\cL_{\lambda,\boldsymbol m}$.  Hence $k\leq K_{\lambda\boldsymbol m}\leq k_{\max}(\boldsymbol m)$.  The gauge-fixed construction \eqref{eq:gauge-fixed-KL} attains the largest multiplicity.  Finally, fixing a gauge vector in any correctable subsystem gives an ordinary correctable subspace of the same logical dimension, to which the bound applies.
\end{proof}

\begin{theorem}[Cyclic monodromy code theorem]\label{thm:cyclic-code}
Fix $\lambda$ with $K_{\lambda\boldsymbol m}>0$.
\begin{enumerate}
\item The factor $\cL_{\lambda,\boldsymbol m}$ in \eqref{eq:young-rule} is a noiseless logical subsystem for every monodromy channel.
\item For every unit vector $v\in S^\lambda$, the gauge-fixed subspace
\begin{equation}\label{eq:cyclic-code-subspace}
 \cC_{\lambda,\boldsymbol m;v}
 =\C v\otimes\cL_{\lambda,\boldsymbol m}
 \subseteq\cH_{\gamma,\boldsymbol m}
\end{equation}
is an exact subspace code of dimension $K_{\lambda\boldsymbol m}$ for the entire error algebra $\cA_{\mathrm{mon}}(\boldsymbol m)$.
\item A recovery may be chosen independently of the particular monodromy channel: on the $\lambda$-isotypic block, trace out the Specht factor and replace it by $\proj{v}$.
\end{enumerate}
\end{theorem}

\begin{proof}
On the $\lambda$-isotypic block, every $A\in\cA_{\mathrm{mon}}(\boldsymbol m)$ has the form
\begin{equation}\label{eq:monodromy-block}
 A\big|_\lambda=A_\lambda\otimes I_{\cL}.
\end{equation}
This proves the noiseless-subsystem statement.  If $P_v=\proj{v}\otimes I_{\cL}$ and $A,B\in\cA_{\mathrm{mon}}$, then
\[
 P_vA^\dagger BP_v
 =\bra{v}A_\lambda^\dagger B_\lambda\ket{v}\,P_v.
\]
The Knill--Laflamme equations therefore hold for any finite family in the algebra and, equivalently, for its whole linear span.

For the recovery, let $W_\lambda:S^\lambda\otimes\cL_{\lambda,\boldsymbol m}\to\cH_{\gamma,\boldsymbol m}$ be the isometric inclusion.  On the isotypic block define
\begin{equation}\label{eq:reset-recovery}
 \cR_{\lambda,v}(\rho)
 =W_\lambda\left(
   \proj{v}\otimes
   \Tr_{S^\lambda}(W_\lambda^\dagger\rho W_\lambda)
   \right)W_\lambda^\dagger.
\end{equation}
To specify a completely positive trace-preserving extension, put
$P_\lambda=W_\lambda W_\lambda^\dagger$, choose any density operator
$\omega_0$ on $\cH_{\gamma,\boldsymbol m}$, and add the term
$\Tr((I-P_\lambda)\rho)\omega_0$ to the right-hand side of
\eqref{eq:reset-recovery}.  A monodromy channel changes only the Specht
density operator in a state initially supported on the code, so
\eqref{eq:reset-recovery} restores the gauge and leaves the logical density
operator unchanged; the added term then vanishes.
\end{proof}

\begin{remark}[What is protected]\label{rem:error-scope}
The theorem corrects a correlated geometric error algebra, not all errors of bounded tensor weight.  It protects against arbitrary root-permutation monodromy inside a fixed regular cyclic type.  Perturbations that change the degrees $m_i$, leave the regular locus, mix different $\lambda$-blocks, or act through the commutant can damage the logical subsystem.  The error model is broad within its geometric class and narrow outside it.
\end{remark}

\subsection{Collision and Young--Jucys--Murphy footprints}

For a transposition $\tau=(ab)$, define
\begin{equation}\label{eq:collision-projectors}
 \Pi_{ab}^{\pm}=\frac12\big(I\pm\rho_{\boldsymbol m}(\tau)\big).
\end{equation}
These are the orthogonal projectors onto the symmetric and antisymmetric subspaces for the local collision monodromy.  They differ from the diagonal projector onto allocations in which $a$ and $b$ have equal colour: the $+$ sector also contains symmetric combinations of two different-colour sheets.  The distinction is forced by quantum coherence.

\begin{proposition}[Collision projectors are logically safe]\label{prop:collision-safe}
On every block $S^\lambda\otimes\cL_{\lambda,\boldsymbol m}$,
\begin{equation}\label{eq:collision-block}
 \Pi_{ab}^{\pm}=\Pi_{ab,\lambda}^{\pm}\otimes I_{\cL}.
\end{equation}
Consequently, for an encoded state $\sigma_G\otimes\rho_L$ with gauge state $\sigma_G$ independent of $\rho_L$, collision-parity measurement reveals no information about $\rho_L$.  If the gauge-fixed code uses a $\pm$ eigenvector of $\rho_\lambda(\tau)$, the measurement is no-error safe.
\end{proposition}

\begin{proof}
The transposition belongs to $S_M$, so its representation has the block form \eqref{eq:monodromy-block}.  Polynomial functional calculus preserves this form and yields \eqref{eq:collision-block}.  Outcome probabilities depend on the gauge density operator only.  If $v$ is an eigenvector, every no-error code state lies in its fixed outcome sector.
\end{proof}

\begin{remark}[A classical colour test need not preserve the code block]\label{rem:classical-colour-test}
Consider $\boldsymbol m=(2,1,1,0)$, realized by $r=4$, $t=1$ and node degrees $(0,1,1,1)$.  In its twelve-word sheet space, let $D_{12}\ket w=\mathbf1_{\{w_1=w_2\}}\ket w$, and let $P_{(3,1)}$ be the central isotypic projector.  The standard character $\chi^{(3,1)}(\sigma)=\#\operatorname{Fix}(\sigma)-1$ and \eqref{eq:central-isotypic-projector} give
\[
 \bra{1213}P_{(3,1)}D_{12}\ket{1123}=\frac18,
 \qquad
 \bra{1213}D_{12}P_{(3,1)}\ket{1123}=0.
\]
For the first matrix element, the two permutations carrying $1123$ to $1213$ are $(23)$ and $(132)$, whose character values are $1$ and $0$; the prefactor is $3/24$.  Thus $D_{12}$ does not preserve this six-dimensional isotypic block.  Classical ramification data and gauge-only collision-parity observables are different measurements.
\end{remark}

The dimensions of the two footprint sectors are also explicit.  Put
\begin{equation}\label{eq:fixed-allocation-count}
 F_{\boldsymbol m}
 =\sum_{i:m_i\geq2}
   \frac{(M-2)!}{(m_i-2)!\prod_{j\neq i}m_j!}
 =\eta(\boldsymbol m)
   \frac{\sum_i m_i(m_i-1)}{M(M-1)}.
\end{equation}

\begin{proposition}[Collision-sector ranks]\label{prop:collision-ranks}
Assume $M\geq2$.  For every transposition $(ab)$,
\begin{equation}\label{eq:full-collision-ranks}
 \operatorname{rank}\Pi_{ab}^{\pm}
 =\frac12\big(\eta(\boldsymbol m)\pm F_{\boldsymbol m}\big).
\end{equation}
On the Specht gauge factor $S^\lambda$, define
\begin{equation}\label{eq:kappa-lambda}
 \kappa_\lambda
 =\sum_{i\geq1}\lambda_i(\lambda_i-2i+1).
\end{equation}
Then the $\pm$ collision eigenspaces have dimensions
\begin{equation}\label{eq:specht-collision-ranks}
 d_\lambda^\pm
 =\frac{f^\lambda}{2}
  \left(1\pm\frac{\kappa_\lambda}{M(M-1)}\right),
\end{equation}
and their ranks on the full $\lambda$-isotypic block are
$K_{\lambda\boldsymbol m}d_\lambda^\pm$.
\end{proposition}

\begin{proof}
The trace of a permutation matrix is the number of fixed basis elements.  An allocation is fixed by $(ab)$ exactly when the two distinguished roots lie in the same colour block.  Choosing that colour and allocating the remaining $M-2$ roots gives the first expression in \eqref{eq:fixed-allocation-count}; division by \eqref{eq:eta} gives the second.  Since $\rho( (ab))$ is an involution, the dimensions of its $\pm1$ eigenspaces are one half of its dimension plus or minus its trace, proving \eqref{eq:full-collision-ranks}.

On $S^\lambda$, the trace is the character value $\chi^\lambda((12))$.  The central class sum $\sum_{a<b}(ab)$ acts on $S^\lambda$ by the scalar $\kappa_\lambda/2$.  Taking traces and using that the transposition class has $M(M-1)/2$ elements gives
\[
 \frac{\chi^\lambda((12))}{f^\lambda}
 =\frac{\kappa_\lambda}{M(M-1)}.
\]
The same involution argument yields \eqref{eq:specht-collision-ranks}; tensoring with the Kostka factor multiplies each rank by its dimension.
\end{proof}

If $(ab)$ and $(cd)$ are disjoint, their collision projectors commute.  If two transpositions share one root, they generally fail to commute.  Thus a set of disjoint collisions may be measured in one commuting round, while overlapping collision tests form different diagnostic contexts.  This is directly analogous to compatible and overlapping charge clusters in categorical footprint models.

\paragraph{Young--Jucys--Murphy footprints.}

The Young--Jucys--Murphy elements are \cite{VershikOkounkov2005}
\begin{equation}\label{eq:YJM}
 J_1=0,
 \qquad
 J_k=\sum_{i=1}^{k-1}(ik)\in\C[S_M],
 \quad 2\leq k\leq M.
\end{equation}
The elements $J_1,\ldots,J_M$ commute.  On a Specht module $S^\lambda$, they have a joint orthonormal eigenbasis $\{v_T:T\in\std(\lambda)\}$ indexed by standard Young tableaux, with
\begin{equation}\label{eq:YJM-content}
 J_kv_T=c_T(k)v_T,
 \qquad
 c_T(k)=\operatorname{column}_T(k)-\operatorname{row}_T(k).
\end{equation}
Let $Q_T^\lambda$ denote the joint spectral projector onto $\C v_T$.

\begin{theorem}[Tableau footprint algebra]\label{thm:YJM-footprint}
Fix $\lambda$ and a tableau $T_0\in\std(\lambda)$.  On the code
\[
 \cC_{\lambda,\boldsymbol m;T_0}
 =\C v_{T_0}\otimes\cL_{\lambda,\boldsymbol m},
\]
the algebra
\begin{equation}\label{eq:tableau-footprint-algebra}
 \cF_\lambda
 =\Span\{Q_T^\lambda\otimes I_{\cL}:T\in\std(\lambda)\}
\end{equation}
is a no-error-safe, logically non-leaking footprint algebra on the selected isotypic block.  Adjoining $I-P_\lambda$ extends its measurement to the full Hilbert space.  It is syndrome-admissible for any chosen error representatives $E_a\in\cA_{\mathrm{mon}}$ for which the nonzero vectors $E_av_{T_0}$ lie in specified, mutually orthogonal joint eigenspaces.
\end{theorem}

\begin{proof}
The no-error code is contained in the $T_0$ sector.  Every joint spectral projector acts as a gauge projector tensored with the identity on the multiplicity space, so its statistics are independent of the logical state.  Under the stated resolution hypothesis, each representative maps the code into its assigned orthogonal sector.  Hence the algebra meets Definition~\ref{def:footprint-algebra}.
\end{proof}

The Young--Jucys--Murphy measurement is not asserted to identify every permutation.  Its virtue is canonical commutativity and logical safety.  The full code remains correctable even when two monodromies have the same tableau footprint, because their residual difference still belongs to the gauge algebra.  This will cease to be true after commutant errors are admitted.

\section{The argyle master space and the cyclic weight sectors}\label{sec:argyle-master}

\subsection{A stable argyle chain with product moduli}

We now construct a compact moduli space whose quantization contains all of the cyclic sheet modules for fixed $M$ and $r$.  The construction uses a rank-one member of the argyle family.  Higher-rank blocks and their Grassmannian factors are treated in Section~\ref{sec:higher-argyle}.

Fix integers $M\geq1$ and $r\geq2$.  Consider the $A_{2M+1}$ quiver with twist
\[
 \widetilde L=\cO_{\Pp^1}(r)
\]
and line-bundle nodes $V_j=\cO_{\Pp^1}(e_j)$, where
\begin{equation}\label{eq:master-degrees}
 \begin{aligned}
 e_{2a-1}&=-(a-1)(r+1),\\
 e_{2a}&=-(a-1)(r+1)-1,
 \end{aligned}
 \qquad 1\leq a\leq M,
 \qquad
 e_{2M+1}=-M(r+1).
\end{equation}
The arrow-section degrees alternate:
\begin{equation}\label{eq:master-arrow-degrees}
 e_{2a}-e_{2a-1}+r=r-1,
 \qquad
 e_{2a+1}-e_{2a}+r=0.
\end{equation}
The quiver has the schematic form
\begin{equation}\label{eq:master-quiver-figure}
\begin{tikzpicture}[baseline=-2pt,>=Latex,node distance=1.17cm]
\node (v1) {$1$};
\node (v2) [right=of v1] {$1$};
\node (v3) [right=of v2] {$1$};
\node (v4) [right=of v3] {$1$};
\node (vd) [right=of v4] {$\cdots$};
\node (ve) [right=of vd] {$1$};
\draw[->] (v1)--node[above] {$r-1$}(v2);
\draw[->] (v2)--node[above] {$0$}(v3);
\draw[->] (v3)--node[above] {$r-1$}(v4);
\draw[->] (v4)--node[above] {$0$}(vd);
\draw[->] (vd)--node[above] {$0$}(ve);
\node[below=5pt of v3,align=center,font=\small] {arrow labels indicate\\section degrees};
\end{tikzpicture}
\end{equation}
and is an argyle quiver with every alternating rank specialized to $1$.

\begin{proposition}[Argyle master space]\label{prop:argyle-master-space}
Every representation of the quiver \eqref{eq:master-quiver-figure} with nonzero arrows is stable for the Higgs-slope stability condition, and its moduli space is
\begin{equation}\label{eq:master-space}
 X_{M,r}\cong(\Pp^{r-1})^M.
\end{equation}
\end{proposition}

\begin{proof}
Over $\C(\Pp^1)$, a chain of line-bundle nodes with nonzero arrows is a single regular nilpotent Jordan block.  Its invariant subspaces are precisely the terminal subspaces.  Their saturations in the split bundle are the corresponding terminal subchains, so these are all saturated Higgs-invariant subbundles.  The degrees \eqref{eq:master-degrees} are strictly decreasing.  The average degree of any proper terminal subsequence is therefore strictly less than the average of the full sequence, so every proper invariant subrepresentation has smaller slope.  This proves stability.

Conversely, if the arrow from $V_j$ to $V_{j+1}$ vanishes, the initial subchain $V_1\oplus\cdots\oplus V_j$ is invariant.  Its average degree is strictly greater than that of the full decreasing sequence, so it destabilizes.  Thus the stable locus is exactly the locus in which every arrow is nonzero.

At an odd arrow, the nonzero map is a nonzero section of $\cO(r-1)$ and contributes a point of
\[
 \Pp H^0(\Pp^1,\cO(r-1))\cong\Pp^{r-1}.
\]
At an even arrow, the nonzero map is a constant and contributes a point.  Quotienting by the line-bundle automorphisms projectivizes the arrow spaces independently: the incidence matrix of a chain has full rank after the global scalar is removed.  Taking the product over the $M$ nontrivial arrows gives \eqref{eq:master-space}.
\end{proof}

The product factors may be permuted, giving an $S_M$-action on $X_{M,r}$.  The diagonal action of $\mathrm{GL}_r$ on $\Pp^{r-1}$ commutes with it.  Although the original labelled chain orders the factors, the product identification equips its moduli with this additional master-space symmetry.

\subsection{Quantization, Schur--Weyl duality, and collision loci}

Let
\begin{equation}\label{eq:master-polarization}
 \mathscr L_{M,r}=\cO_{X_{M,r}}(1,\ldots,1).
\end{equation}
Fix an identification $H^0(\Pp^1,\cO(r-1))\cong\C^r$, including an ordered orthonormal coefficient basis.  It fixes the diagonal torus, the Fubini--Study structures and their normalized section inner products.  Define the quantization
\begin{equation}\label{eq:master-Hilbert}
 \cK_{M,r}=H^0(X_{M,r},\mathscr L_{M,r})^*.
\end{equation}
K\"unneth's theorem gives a unitary identification
\begin{equation}\label{eq:master-tensor}
 \cK_{M,r}\cong(\C^r)^{\otimes M}.
\end{equation}
The $S_M$-action permutes tensor factors, while $U(r)$ acts diagonally.  Schur--Weyl duality \cite{FultonHarris1991} yields
\begin{equation}\label{eq:schur-weyl}
 \cK_{M,r}
 \cong
 \bigoplus_{\substack{\lambda\vdash M\\\ell(\lambda)\leq r}}
 S^\lambda\otimes \mathbb S_\lambda(\C^r),
\end{equation}
where $\mathbb S_\lambda(\C^r)$ is the irreducible polynomial $\mathrm{GL}_r$-module of highest weight $\lambda$.

Let $T^r\subset U(r)$ be the diagonal torus.  For a composition $\boldsymbol m=(m_1,\ldots,m_r)$ of $M$, write $\cK_{M,r}[\boldsymbol m]$ for the weight space on which
\[
 \diag(z_1,\ldots,z_r)
\]
acts by $z_1^{m_1}\cdots z_r^{m_r}$.  If $e_1,\ldots,e_r$ is the standard basis of $\C^r$, then this weight space has orthonormal word basis
\begin{equation}\label{eq:word-basis}
 e_{c_1}\otimes\cdots\otimes e_{c_M},
 \qquad
 \#\{a:c_a=i\}=m_i.
\end{equation}

\begin{theorem}[Argyle--cyclic weight-sector bridge]\label{thm:argyle-cyclic-bridge}
Let $M=rt$, let $Q$ be a type $(1,\ldots,1)$ cyclic quiver with arrow degrees $\boldsymbol m$, and let $\gamma\in\cB_M^{\reg}$.  Relative to the fixed coefficient basis and an ordering of the zeroes of $\gamma$, there is an $S_M$-equivariant unitary isomorphism
\begin{equation}\label{eq:bridge-isomorphism}
 \Theta_{\gamma,\boldsymbol m}:
 \cH_{\gamma,\boldsymbol m}
 \xrightarrow{\ \cong\ }
 \cK_{M,r}[\boldsymbol m].
\end{equation}
Under this isomorphism, Young's rule \eqref{eq:young-rule} is the $\boldsymbol m$-weight restriction of Schur--Weyl duality:
\begin{equation}\label{eq:weight-schur-weyl}
 \cK_{M,r}[\boldsymbol m]
 \cong
 \bigoplus_{\substack{\lambda\vdash M\\\ell(\lambda)\leq r}}
 S^\lambda\otimes\mathbb S_\lambda(\C^r)[\boldsymbol m],
 \qquad
 \dim\mathbb S_\lambda(\C^r)[\boldsymbol m]
 =K_{\lambda\boldsymbol m}.
\end{equation}
\end{theorem}

\begin{proof}
Given an allocation $(B_1,\ldots,B_r)$, define a word $c_1\cdots c_M$ by $c_a=i$ when $z_a\in B_i$.  The corresponding vector \eqref{eq:word-basis} has content $\boldsymbol m$.  This is a bijection between orthonormal bases and hence a unitary isomorphism.  A root permutation acts on an allocation and on its word by permuting positions, so the map is $S_M$-equivariant.

Taking the $\boldsymbol m$-weight subspace of \eqref{eq:schur-weyl} gives the first part of \eqref{eq:weight-schur-weyl}.  The weight multiplicity in the Schur module is the Kostka number by the semistandard-tableau character formula.  This is precisely the multiplicity in Young's rule.
\end{proof}

The root ordering used in \Cref{thm:argyle-cyclic-bridge} can be removed by descent; the chosen coefficient basis and tensor-position action remain part of the master-space construction.  To make this precise, let
\[
 \operatorname{Conf}_M(\Pp^1)
 =\{(z_1,\ldots,z_M):(z_a\neq z_b)\text{ for }a\neq b\}
\]
and let $\operatorname{UConf}_M(\Pp^1)=\operatorname{Conf}_M(\Pp^1)/S_M$.  The latter is naturally the projectivized regular cyclic base.  Form the associated unitary local systems
\begin{align}
 \mathscr H_{\boldsymbol m}
 &=\operatorname{Conf}_M(\Pp^1)
   \times_{S_M}\C[S_M/S_{\boldsymbol m}],
   \label{eq:cyclic-local-system}\\
 \mathscr K_{\boldsymbol m}
 &=\operatorname{Conf}_M(\Pp^1)
   \times_{S_M}\cK_{M,r}[\boldsymbol m].
   \label{eq:argyle-local-system}
\end{align}

\begin{corollary}[Global argyle--cyclic bridge]\label{cor:global-bridge}
For the fixed master-space choices, the maps $\Theta_{\gamma,\boldsymbol m}$ descend to a unitary isomorphism of local systems
\begin{equation}\label{eq:global-bridge}
 \mathscr H_{\boldsymbol m}\cong\mathscr K_{\boldsymbol m}
 \qquad\text{over }\operatorname{UConf}_M(\Pp^1).
\end{equation}
Under \eqref{eq:global-bridge}, parallel transport is the cyclic-cover monodromy on the left and tensor-position permutation on the right.  In particular, the bridge is independent of a global ordering of the spectral zeroes.
\end{corollary}

\begin{proof}
On the ordered configuration cover, both bundles are constant and the word-allocation map in the proof of \Cref{thm:argyle-cyclic-bridge} is $S_M$-equivariant.  It therefore descends through the free $S_M$-quotient.  The associated-bundle construction identifies deck transformations with the two stated monodromy actions.
\end{proof}

The system $\mathscr K_{\boldsymbol m}$ is associated to the root-ordering cover using the chosen tensor-permutation symmetry of the fixed argyle quantization.  The corollary does not identify it with a Gauss--Manin system of an independently varying argyle family, or derive factor swaps from loops around argyle resultant divisors.

\begin{figure}[t]
\centering
\begin{tikzpicture}[>=Latex,node distance=1.35cm and 1.8cm]
\node[draw,rounded corners,fill=blue!5,align=center] (cyc) {regular cyclic fibre\\$h_Q^{-1}(\gamma)$};
\node[draw,rounded corners,fill=blue!5,align=center,right=of cyc] (alloc) {root allocations\\$\Omega_{\boldsymbol m}$};
\node[draw,rounded corners,fill=green!7,align=center,right=3cm of alloc] (weight) {weight-sector basis\\$\cK_{M,r}[\boldsymbol m]$};
\node[draw,rounded corners,fill=orange!9,align=center,below=of weight] (sw) {Schur--Weyl block\\$S^\lambda\otimes\mathbb S_\lambda(\C^r)[\boldsymbol m]$};
\draw[<->] (cyc)--node[above,font=\small] {divisors}(alloc);
\draw[<->] (alloc)--node[above,font=\small] {coloured words}(weight);
\draw[->] (weight)--node[right,font=\small] {isotypic projection}(sw);
\node[draw,dashed,rounded corners,fit=(cyc)(alloc)(weight),label=below left:{\small $S_M$ permutes roots/positions}] {};
\end{tikzpicture}
\caption{Linearizing the sheet-to-word bijection identifies the cyclic sheet Hilbert space with a torus-weight sector of the quantized argyle master space.  Monodromy acts on the Specht factor; the Kostka weight space is logical.}
\label{fig:bridge}
\end{figure}
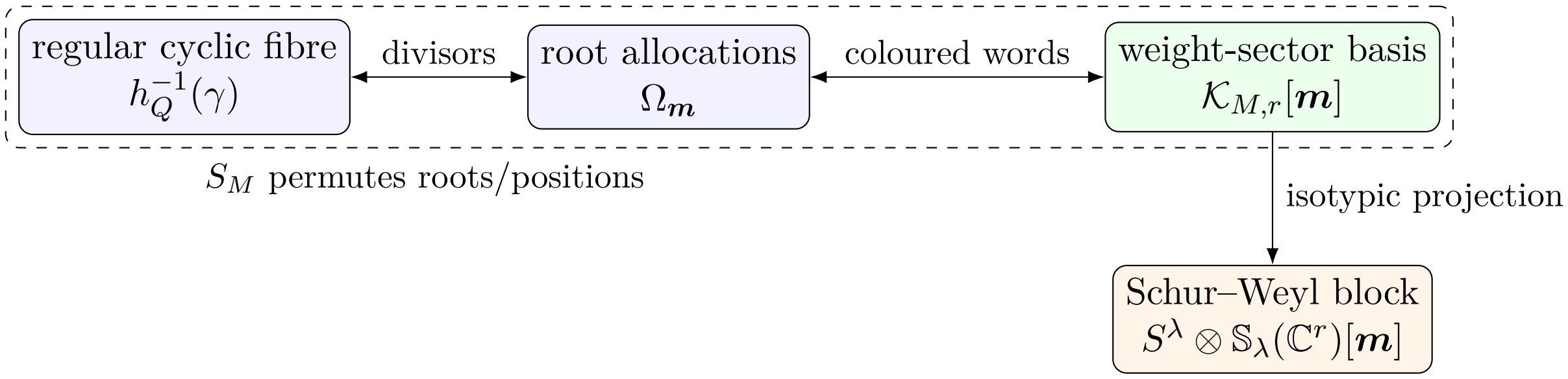

\begin{corollary}[Master-space code]\label{cor:master-code}
For any $\lambda\vdash M$ with $\ell(\lambda)\leq r$, the Schur module $\mathbb S_\lambda(\C^r)$ is a noiseless subsystem of $\cK_{M,r}$ for the tensor-factor permutation algebra $\C[S_M]$.  Fixing a vector in $S^\lambda$ gives an exact subspace code of dimension $\dim\mathbb S_\lambda(\C^r)$.  Its intersection with the torus-weight-$\boldsymbol m$ sector is the cyclic code of \Cref{thm:cyclic-code} and has dimension $K_{\lambda\boldsymbol m}$.
\end{corollary}

\begin{remark}[Dual roles of the commuting actions]\label{rem:dual-actions}
The two commuting actions have operationally different roles.  When root permutations are the errors, the Specht factor is gauge and the $\mathrm{GL}_r$ factor is logical.  If instead collective $U(r)$ rotations were the noise, the roles would reverse and Specht multiplicities would be protected.  We use the first interpretation because it is selected by cyclic-cover monodromy.
\end{remark}

\paragraph{Collision loci in the master space.}

Write a point of the $a$th factor $\Pp^{r-1}$ as a degree-$(r-1)$ binary form $p_a$.  The hypersurface
\begin{equation}\label{eq:resultant-divisor}
 D_{ab}=\{\Res(p_a,p_b)=0\}\subset X_{M,r}
\end{equation}
records a common polynomial zero between two argyle blocks.  These resultantal hypersurfaces are rank-one analogues of the invariant-zero collision manifolds in \cite{RayanSundboArgyle2018}.  The diagonal $p_a=p_b$ is contained in $D_{ab}$ and is the fixed locus of the factor transposition $(ab)$.

Quantization sends factor transposition to the swap operator on the $a$th and $b$th copies of $\C^r$.  On the cyclic weight sector, this is exactly the root transposition in \eqref{eq:collision-projectors}.  Thus the master-space symmetry and cyclic collision monodromy define the same operator on the shared weight sector.  This operator identification does not assert that a loop around $D_{ab}$ produces a factor swap in the argyle moduli space.  For $r>2$, the full resultant divisor is richer: it detects a shared zero without requiring equality of the two polynomial factors.  When $r=2$, the forms are linear and $D_{ab}$ equals the diagonal.  Section~\ref{sec:higher-argyle} states the additional measurement and error-operator assumptions under which such classical collision data satisfy an exact recovery criterion.

\section{Explicit cyclic and argyle code families}\label{sec:six-sheet}

\subsection{The six-sheet quiver code: geometry and logical states}

Take $r=3$, $t=1$, and $U_1=U_2=U_3=\cO_{\Pp^1}$.  A representation is
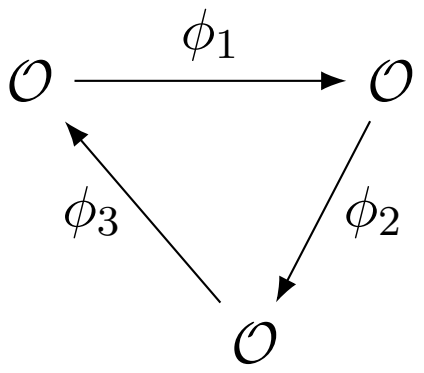
\begin{equation}\label{eq:three-cycle}
\begin{tikzpicture}[baseline=-2pt,>=Latex,node distance=1.8cm]
\node (u1) {$\cO$};
\node (u2) [right=of u1] {$\cO$};
\node (u3) [below right=1.2cm and .9cm of u1] {$\cO$};
\draw[->] (u1)--node[above] {$\phi_1$}(u2);
\draw[->] (u2)--node[right] {$\phi_2$}(u3);
\draw[->] (u3)--node[left] {$\phi_3$}(u1);
\end{tikzpicture}
\qquad
 \phi_i\in H^0(\Pp^1,\cO(1)).
\end{equation}
For a generic cubic
\[
 \gamma=c\,(z-z_1)(z-z_2)(z-z_3),
\]
each arrow receives one root.  A sheet is therefore a bijection between the three roots and the three arrow colours.  After fixing a reference bijection, the six sheets are identified with $S_3$ and
\begin{equation}\label{eq:S3-regular}
 \cH_{\gamma,(1,1,1)}\cong\C[S_3]
\end{equation}
with root monodromy acting by the left regular representation.  Keeping the node labels is essential: cyclically rotating three equal line-bundle summands conjugates the ungraded Higgs matrix and can identify distinct allocations after forgetting the grading.

\begin{figure}[t]
\centering
\begin{tikzpicture}[>=Latex,scale=1.05,every node/.style={font=\small}]
\node[draw,circle,fill=blue!6] (e) at (90:2.25) {$e$};
\node[draw,circle,fill=blue!6] (a) at (30:2.25) {$s_1$};
\node[draw,circle,fill=blue!6] (b) at (-30:2.25) {$s_2s_1$};
\node[draw,circle,fill=blue!6] (c) at (-90:2.25) {$s_1s_2s_1$};
\node[draw,circle,fill=blue!6] (d) at (-150:2.25) {$s_1s_2$};
\node[draw,circle,fill=blue!6] (f) at (150:2.25) {$s_2$};
\draw[very thick,red!65!black] (e)--node[above right] {$s_1$}(a);
\draw[very thick,blue!65!black] (a)--node[right] {$s_2$}(b);
\draw[very thick,red!65!black] (b)--node[below right] {$s_1$}(c);
\draw[very thick,blue!65!black] (c)--node[below left] {$s_2$}(d);
\draw[very thick,red!65!black] (d)--node[left] {$s_1$}(f);
\draw[very thick,blue!65!black] (f)--node[above left] {$s_2$}(e);
\node[align=center] at (0,0) {six root\\allocations};
\end{tikzpicture}
\caption{The six cyclic sheets form the $S_3$ Cayley graph.  The generators $s_1=(12)$ and $s_2=(23)$ are simple root-collision monodromies.}
\label{fig:S3-sheets}
\end{figure}
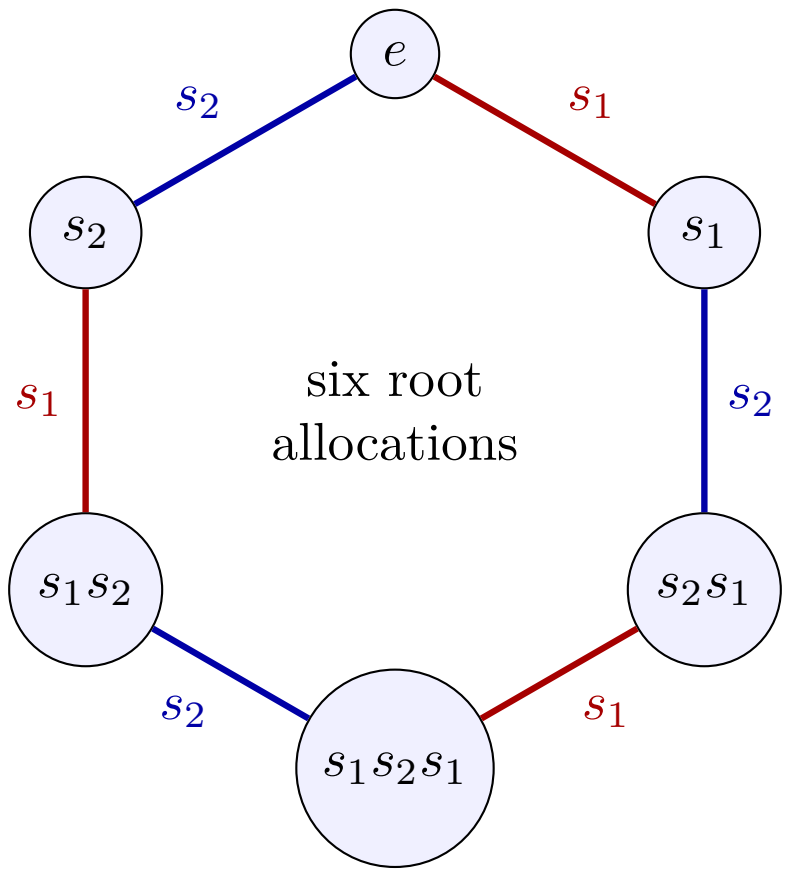

The regular representation decomposes as
\begin{equation}\label{eq:S3-decomp}
 \C[S_3]
 \cong
 \mathbf1\oplus
 \big(S^{(2,1)}\otimes\C^2\big)
 \oplus\sgn.
\end{equation}
The middle summand has a two-dimensional Specht gauge factor and a two-dimensional logical multiplicity factor.

The balance of the three arrow degrees is essential in this smallest case.  Up to relabelling colours, the three compositions of $3$ give the modules listed in \Cref{tab:S3-compositions}.
\begin{table}[H]
\centering
\small
\renewcommand{\arraystretch}{1}
\begin{tabular}{@{}>{\raggedright\arraybackslash}p{0.15\textwidth}@{\hspace{1.5em}}>{\raggedright\arraybackslash}p{0.44\textwidth}@{\hspace{1.5em}}>{\raggedright\arraybackslash}p{\dimexpr0.34\textwidth-1.5em\relax}@{}}
$\boldsymbol m$ & \textbf{Cyclic sheet module} & \textbf{Largest Kostka multiplicity}\\[3pt]
\hline
\noalign{\vskip 5pt}
$(3,0,0)$ & $\mathbf1$ & $1$\\[4pt]
$(2,1,0)$ & $\mathbf1\oplus S^{(2,1)}$ & $1$\\[4pt]
$(1,1,1)$ & $\mathbf1\oplus(S^{(2,1)}\otimes\C^2)\oplus\sgn$ & $2$\\
\end{tabular}
\caption{Cyclic sheet modules for the three compositions of $3$, up to colour relabelling.}
\label{tab:S3-compositions}
\end{table}
Thus a three-sheet permutation representation does not yet encode a logical system: the logical qubit appears only when all six ordered root-to-arrow allocations are present.

\paragraph{Explicit logical states.}

Let $s_1=(12)$ and $s_2=(23)$.  Use the real standard representation
\begin{equation}\label{eq:S3-standard-matrices}
 D(s_1)=
 \begin{pmatrix}1&0\\0&-1\end{pmatrix},
 \qquad
 D(s_2)=
 \begin{pmatrix}-\frac12&\frac{\sqrt3}{2}\\[2pt]
 \frac{\sqrt3}{2}&\frac12\end{pmatrix}.
\end{equation}
The nonabelian Fourier vectors
\begin{equation}\label{eq:S3-Fourier}
 \ket{a,b}
 =\frac1{\sqrt3}\sum_{g\in S_3}D(g)_{ab}\ket{g},
 \qquad a,b\in\{1,2\},
\end{equation}
form an orthonormal basis for the standard isotypic component.  Left multiplication changes $a$ and leaves $b$ unchanged.  We fix $a=1$ and define
\begin{equation}\label{eq:S3-codewords-abstract}
 \ket{0_L}=\ket{1,1},
 \qquad
 \ket{1_L}=\ket{1,2}.
\end{equation}
In the sheet basis these are
\begin{align}
 \ket{0_L}
 &=\frac1{\sqrt3}\left(
 \ket e+\ket{s_1}
 -\frac12\big(
 \ket{s_2}+\ket{s_1s_2}+\ket{s_2s_1}+\ket{s_1s_2s_1}
 \big)\right),\label{eq:S3-codeword0}\\
 \ket{1_L}
 &=\frac12\left(
 \ket{s_2}+\ket{s_1s_2}
 -\ket{s_2s_1}-\ket{s_1s_2s_1}
 \right).\label{eq:S3-codeword1}
\end{align}

\begin{proposition}[Exact six-sheet qubit]\label{prop:six-sheet-code}
The span of \eqref{eq:S3-codeword0}--\eqref{eq:S3-codeword1} is an exact one-qubit code for every channel whose Kraus operators are linear combinations of left root permutations.  An explicit recovery performs the $S_3$ Fourier transform, discards the gauge index $a$, resets it to $1$, and applies the inverse transform.
\end{proposition}

\begin{proof}
This is \Cref{thm:cyclic-code} for $\lambda=(2,1)$ and $K_{(2,1),(1,1,1)}=2$.  In the basis \eqref{eq:S3-Fourier}, left multiplication has the form $D(g)$ on the $a$ register tensored with the identity on the $b$ register.  Indeed, real orthogonality gives $L_h\ket{a,b}=\sum_c D(h)_{ca}\ket{c,b}$; similarly $R_h\ket{a,b}=\sum_c D(h)_{cb}\ket{a,c}$.  The stated recovery is \eqref{eq:reset-recovery} in Fourier coordinates.
\end{proof}

\subsection{Argyle realization and collision footprints}

For $M=r=3$, the master quiver of \Cref{prop:argyle-master-space} has twist $\cO(3)$, node degrees
\begin{equation}\label{eq:A7-degrees}
 0,-1,-4,-5,-8,-9,-12,
\end{equation}
and arrow-section degrees
\[
 2,0,2,0,2,0.
\]
It is a stable $A_7$ argyle chain with
\begin{equation}\label{eq:A7-master}
 X_{3,3}\cong(\Pp^2)^3,
 \qquad
 H^0(X_{3,3},\cO(1,1,1))^*\cong(\C^3)^{\otimes3}.
\end{equation}
The $(1,1,1)$ torus-weight sector is the span of the six words
\[
 e_{\sigma(1)}\otimes e_{\sigma(2)}\otimes e_{\sigma(3)},
 \qquad \sigma\in S_3,
\]
and \Cref{thm:argyle-cyclic-bridge} identifies it with \eqref{eq:S3-regular}.  Thus the six-sheet qubit is simultaneously a cyclic Hitchin-fibre code and a weight-sector code in a three-qutrit argyle quantization.

\paragraph{Collision footprints and an invisible logical error.}

In the standard tableau basis, $J_2=s_1$ acts as $D(s_1)$ on the gauge register.  The code fixes the $+1$ gauge eigenvector, so the measurement
\begin{equation}\label{eq:S3-footprint-projectors}
 \Pi_+=\frac12(I+L_{s_1}),
 \qquad
 \Pi_-=\frac12(I-L_{s_1}),
\end{equation}
is no-error safe and reveals no logical information.  More generally, every observable in the left group algebra acts only on the gauge register.

To see the limitation, let $L_g$ and $R_g$ denote left and right regular actions, with the convention $R_h\ket g=\ket{gh^{-1}}$.  These actions commute.  On the standard isotypic block, $R_h$ acts on the logical index $b$.  Consider
\begin{equation}\label{eq:logical-paulis-right-action}
 \overline Z=PR_{s_1}P,
 \qquad
 \overline X
 =P\left(\frac1{\sqrt3}R_{s_1}
          +\frac2{\sqrt3}R_{s_2}\right)P,
 \qquad
 \overline Y=i\overline X\,\overline Z,
\end{equation}
where $P$ is the projector onto the gauge-fixed code.

\begin{proposition}[Complete logical commutant]\label{prop:S3-logical-commutant}
The operators in \eqref{eq:logical-paulis-right-action} are the logical Pauli matrices $Z,X,Y$ in the basis \eqref{eq:S3-codewords-abstract}.  The compression of the right group algebra to the code is the full logical algebra:
\begin{equation}\label{eq:right-full-logical-algebra}
 P\,R(\C[S_3])\,P=\End(\cC)\cong M_2(\C).
\end{equation}
Every logical unitary has a unitary extension in the right algebra that commutes with every root-monodromy operator.
\end{proposition}

\begin{proof}
On the logical index, the two right transpositions act by the real matrices in \eqref{eq:S3-standard-matrices}.  Hence $R_{s_1}$ compresses to $Z$, while
\[
 \frac1{\sqrt3}D(s_1)+\frac2{\sqrt3}D(s_2)
 =\begin{pmatrix}0&1\\1&0\end{pmatrix}=X.
\]
These matrices generate $M_2(\C)$, proving \eqref{eq:right-full-logical-algebra}.  Equivalently, this is the standard block of the commutant formula \eqref{eq:mon-commutant-wedderburn}.  The right algebra has Wedderburn form $\C\oplus M_2(\C)\oplus\C$; adjoining arbitrary phases on its two one-dimensional blocks extends any $U\in U(2)$ to a unitary element of the full right algebra.
\end{proof}

Now consider
\begin{equation}\label{eq:same-footprint-errors}
 E=L_{s_2},
 \qquad
 E'=L_{s_2}R_{s_1}.
\end{equation}
The first is pure root monodromy.  The second is the same root monodromy followed by an abstract permutation of two arrow colours in the sheet register.  This added operation lies outside the monodromy error algebra and is not a bundle-gauge identification.

\begin{theorem}[Same-footprint logical ambiguity]\label{thm:same-footprint-S3}
For every footprint measurement contained in the left monodromy algebra, $E$ and $E'$ in \eqref{eq:same-footprint-errors} have identical outcome probabilities on every encoded state.  Nevertheless,
\begin{equation}\label{eq:S3-residual-logical}
 P E^\dagger E'P
 =P R_{s_1}P
 =\overline Z,
\end{equation}
where $\overline Z$ is the non-scalar logical Pauli $Z$ in the basis \eqref{eq:S3-codewords-abstract}.  Hence the enlarged error family $\{E,E'\}$ is not exactly correctable by any recovery on this code.
\end{theorem}

\begin{proof}
Every left-algebra projector has the form $Q_G\otimes I_L$ on the standard block.  Since $R_{s_1}$ commutes with the left action and changes only the logical factor, inserting it after $E$ cannot change the probability of any such projector.  With the stated Fourier and right-action conventions, the logical action of $s_1$ is exactly $D(s_1)$.  Equation \eqref{eq:S3-standard-matrices} therefore gives $R_{s_1}|_L=\diag(1,-1)=\overline Z$.  The within-footprint Knill--Laflamme compression is non-scalar, so exact correction of the pair is impossible.
\end{proof}

\begin{corollary}[Full same-footprint logical fibre]\label{cor:full-same-footprint-fibre}
Let $U$ be any logical unitary on the six-sheet qubit and let $\widetilde U$ be a commuting unitary extension supplied by \Cref{prop:S3-logical-commutant}.  For every unitary left-monodromy error $E$, the pair $E$ and $E\widetilde U$ has identical statistics for every left-algebra footprint measurement on encoded states, while
\begin{equation}\label{eq:full-residual-unitary}
 PE^\dagger E\widetilde U P=U
\end{equation}
where $U$ is understood as an operator on the code.  The pair is exactly correctable if and only if $U$ is scalar.
\end{corollary}

\begin{proof}
The footprint projectors act on the gauge factor and $\widetilde U$ acts on the logical factor, so all outcome probabilities agree.  Equation \eqref{eq:full-residual-unitary} follows from $E^\dagger E=I$ and the definition of the extension.  The Knill--Laflamme condition is exactly the requirement that this compression be scalar.
\end{proof}

This example separates three notions.  The root history $L_{s_2}$ is latent.  The measured left-algebra sector is its footprint.  The commuting colour relabeling is invisible to that footprint but may be logical.  The construction is the finite quiver analogue of two topological error chains with the same boundary and a nontrivial residual cycle.

\subsection{A balanced three-cycle family}\label{sec:balanced}

The six-sheet code is the first member of a growing family.  Let $Q_t$ be the three-node cyclic quiver with $U_i=\cO_{\Pp^1}$ and twist $\cO_{\Pp^1}(t)$.  We work over regular spectral polynomials; \Cref{lem:regular-cyclic-stability} proves that every allocation is stable even though the rank and total degree are not coprime.  Each arrow polynomial has degree $t$, so
\begin{equation}\label{eq:balanced-composition}
 \boldsymbol m=(t,t,t),
 \qquad M=3t,
\end{equation}
and
\begin{equation}\label{eq:balanced-sheet-dim}
 \dim\cH_{\gamma,(t,t,t)}
 =\frac{(3t)!}{(t!)^3}.
\end{equation}

For $0\leq j\leq\lfloor3t/2\rfloor$, set
\begin{equation}\label{eq:balanced-two-row-shape}
 \lambda_{t,j}=(3t-j,j).
\end{equation}

\begin{theorem}[Balanced two-row Kostka spectrum]\label{thm:balanced-two-row}
Every two-row block in the balanced composition has multiplicity
\begin{equation}\label{eq:balanced-two-row-kostka}
 K_{(3t-j,j),(t,t,t)}
 =
 \begin{cases}
  j+1,&0\leq j\leq t,\\
  3t-2j+1,&t\leq j\leq\lfloor3t/2\rfloor.
 \end{cases}
\end{equation}
Consequently the unique maximizing two-row shape is
\begin{equation}\label{eq:balanced-lambda}
 \lambda_t=\lambda_{t,t}=(2t,t),
\end{equation}
and its multiplicity is
\begin{equation}\label{eq:balanced-kostka}
 K_{(2t,t),(t,t,t)}=t+1.
\end{equation}
\end{theorem}

\begin{proof}
Let $h_q(x,y,z)$ be the complete homogeneous symmetric polynomial, with $h_{-1}=0$.  Jacobi--Trudi gives
\[
 s_{(3t-j,j)}
 =h_{3t-j}h_j-h_{3t-j+1}h_{j-1}.
\]
Let
\[
 N_t(q)=\#\{(a,b,c)\in\{0,\ldots,t\}^3:a+b+c=q\},
 \qquad N_t(-1)=0.
\]
Choosing the monomial contributed by the second complete homogeneous factor shows that
\begin{equation}\label{eq:kostka-first-difference}
 K_{(3t-j,j),(t,t,t)}=N_t(j)-N_t(j-1).
\end{equation}
For $0\leq q\leq t$, stars and bars gives $N_t(q)=\binom{q+2}{2}$.  For $t<q\leq\lfloor3t/2\rfloor$, inclusion--exclusion gives
\[
 N_t(q)=\binom{q+2}{2}-3\binom{q-t+1}{2},
\]
because in this range two coordinates cannot simultaneously exceed $t$.  Taking the first difference in \eqref{eq:kostka-first-difference} yields \eqref{eq:balanced-two-row-kostka}.  The first branch increases strictly to $t+1$, and the second branch decreases strictly, proving the maximizing statement and \eqref{eq:balanced-kostka}.
\end{proof}

\begin{theorem}[Complete balanced spectrum and unique maximal block]\label{thm:balanced-full-spectrum}
For every partition $\lambda=(a,b,c)$ of $3t$ with at most three rows, padding with zero parts when necessary,
\begin{equation}\label{eq:balanced-full-kostka}
 K_{(a,b,c),(t,t,t)}=1+\min(a-b,b-c).
\end{equation}
Partitions with more than three rows have multiplicity zero.  The unique partition maximizing the multiplicity is $(2t,t,0)$, with value $t+1$.
\end{theorem}

\begin{proof}
The determinant-shift identity for Schur polynomials gives
\[
 s_{(a,b,c)}(x,y,z)=(xyz)^c s_{(a-c,b-c)}(x,y,z).
\]
Set $u=t-c$ and $j=b-c$.  Then $u\geq0$, $0\leq j\leq\lfloor3u/2\rfloor$, and $a-c=3u-j$.  If $u=0$ the multiplicity is one.  Otherwise \Cref{thm:balanced-two-row} gives
\[
 K_{(a,b,c),(t,t,t)}
 =K_{(3u-j,j),(u,u,u)}
 =1+\min(j,3u-2j),
\]
which is \eqref{eq:balanced-full-kostka}.  A tableau on three letters has no column of height greater than three, proving the vanishing statement.

Write $p=a-b$, $q=b-c$, and $m=\min(p,q)$.  Since
\[
 3t=3c+p+2q\geq3c+3m,
\]
we have $m\leq t$.  Equality requires $c=0$ and $p=q=t$, hence $(a,b,c)=(2t,t,0)$.  This shape attains the bound.
\end{proof}

The spectrum calculation is the three-variable weight-multiplicity formula in elementary form; its coding consequence is the sharp optimality statement below.  For the maximizing shape, the $t+1$ logical basis vectors have a particularly direct tableau indexing.  If $x\in\{0,\ldots,t\}$, take $x$ height-two columns of type $(1,2)$, followed by $t-x$ columns of type $(1,3)$, and fill the remaining top-row boxes with $t-x$ copies of $2$ followed by $x$ copies of $3$.  These are exactly the semistandard tableaux of shape $(2t,t)$ and content $(t,t,t)$.

More generally, every value greater than one on the right-hand side of \eqref{eq:balanced-two-row-kostka} produces a distinct exact two-row code block.  The choice $j=t$ is therefore not merely convenient: by \Cref{thm:balanced-full-spectrum}, it has the largest logical dimension among all isotypic sectors of the balanced sheet module.

The hook-length formula for the Specht gauge factor gives
\begin{equation}\label{eq:balanced-specht-dim}
 f^{(2t,t)}
 =\frac{t+1}{2t+1}\binom{3t}{t}.
\end{equation}

\begin{theorem}[Optimal balanced cyclic logical qudit]\label{thm:balanced-code}
For every $t\geq1$, the regular cyclic sheet Hilbert space for $Q_t$ contains an isotypic block
\begin{equation}\label{eq:balanced-block}
 S^{(2t,t)}\otimes\C^{t+1}.
\end{equation}
After fixing any unit Specht vector, the second factor is an exact logical qudit of dimension $t+1$ for every monodromy channel generated by permutations of the $3t$ spectral zeroes.  No exact subspace or subsystem code for the full monodromy algebra on this sheet space has larger logical dimension.  Equivalently, the optimal number of whole logical qubits is $\lfloor\log_2(t+1)\rfloor$, attained by choosing a $2^{\lfloor\log_2(t+1)\rfloor}$-dimensional logical subspace.
\end{theorem}

\begin{proof}
Equations \eqref{eq:young-rule} and \eqref{eq:balanced-kostka} give the block \eqref{eq:balanced-block}, and \Cref{thm:cyclic-code} supplies recovery.  Equations \eqref{eq:optimal-dimension} and \eqref{eq:balanced-full-kostka} prove optimality over the entire sheet space.
\end{proof}

\begin{example}[The nine-by-three block at $t=2$]\label{ex:t2-qutrit}
For $t=2$, the maximizing shape is $(4,2)$.  The three semistandard logical labels are listed in \Cref{tab:t2-logical-labels}.
\begin{table}[H]
\centering
\small
\renewcommand{\arraystretch}{1}
\begin{tabular}{@{}>{\raggedright\arraybackslash}p{0.15\textwidth}@{\hspace{1.5em}}>{\raggedright\arraybackslash}p{0.40\textwidth}@{\hspace{1.5em}}>{\raggedright\arraybackslash}p{\dimexpr0.38\textwidth-1.5em\relax}@{}}
$\bm{x}$ & \textbf{Top row} & \textbf{Bottom row}\\[3pt]
\hline
\noalign{\vskip 5pt}
$0$ & $1\;1\;2\;2$ & $3\;3$\\[4pt]
$1$ & $1\;1\;2\;3$ & $2\;3$\\[4pt]
$2$ & $1\;1\;3\;3$ & $2\;2$\\
\end{tabular}
\caption{The semistandard logical labels for the $t=2$ code.}
\label{tab:t2-logical-labels}
\end{table}
The Specht factor has dimension $f^{(4,2)}=9$, so the code block has dimension $27$ inside the $90$-sheet Hilbert space.  Fixing one of the nine standard-tableau gauge states leaves an exact logical qutrit.  For comparison, the complete decomposition is
\begin{align*}
 \cH_{\gamma,(2,2,2)}\cong{}&S^{(6)}
 \oplus(S^{(5,1)}\otimes\C^2)
 \oplus(S^{(4,2)}\otimes\C^3)
 \oplus S^{(4,1,1)}\\
 &\oplus S^{(3,3)}
 \oplus(S^{(3,2,1)}\otimes\C^2)
 \oplus S^{(2,2,2)}.
\end{align*}
The dimensions sum to $1+10+27+10+5+32+5=90$, and no other block has multiplicity three.
\end{example}

\begin{corollary}[Balanced collision ranks]\label{cor:balanced-collision-ranks}
On the gauge factor $S^{(2t,t)}$, every simple collision has $\pm$ eigenspace dimensions
\begin{equation}\label{eq:balanced-collision-ranks}
 d_t^\pm
 =\frac{f^{(2t,t)}}2
  \left(1\pm\frac{5(t-1)}{3(3t-1)}\right).
\end{equation}
Each sector is tensored with the same $(t+1)$-dimensional logical factor.  In particular, the ranks are $(1,1)$ at $t=1$ and $(6,3)$ at $t=2$.
\end{corollary}

\begin{proof}
For $\lambda=(2t,t)$, equation \eqref{eq:kappa-lambda} gives
$\kappa_\lambda=5t(t-1)$.  Substitute this and $M=3t$ into \eqref{eq:specht-collision-ranks}.
\end{proof}

\begin{table}[H]
\centering
\small
\renewcommand{\arraystretch}{1}
\begin{tabular}{@{}>{\raggedright\arraybackslash}p{0.04\textwidth}@{\hspace{1.5em}}>{\raggedright\arraybackslash}p{0.22\textwidth}@{\hspace{1.5em}}>{\raggedright\arraybackslash}p{0.17\textwidth}@{\hspace{1.5em}}>{\raggedright\arraybackslash}p{0.22\textwidth}@{\hspace{1.5em}}>{\raggedright\arraybackslash}p{\dimexpr0.28\textwidth-4.5em\relax}@{}}
$\bm{t}$ & \textbf{Roots / master qutrits} & \textbf{Sheet dimension} & \textbf{Specht gauge dimension} & \textbf{Logical dimension}\\[3pt]
\hline
\noalign{\vskip 5pt}
$1$ & $3$  & $6$      & $2$    & $2$\\[4pt]
$2$ & $6$  & $90$     & $9$    & $3$\\[4pt]
$3$ & $9$  & $1680$   & $48$   & $4$\\[4pt]
$4$ & $12$ & $34650$  & $275$  & $5$\\[4pt]
$5$ & $15$ & $756756$ & $1638$ & $6$\\
\end{tabular}
\caption{The first balanced three-cycle codes.  The indicated isotypic block has dimension $f^{(2t,t)}(t+1)$ inside the full sheet Hilbert space.}
\label{tab:balanced-codes}
\end{table}

The corresponding argyle master space is
\begin{equation}\label{eq:balanced-master}
 X_{3t,3}=(\Pp^2)^{3t},
 \qquad
 \cK_{3t,3}=(\C^3)^{\otimes3t}.
\end{equation}
Thus $Q_t$ is the balanced $(t,t,t)$ weight sector of a $3t$-qutrit quantization.  The logical basis following \Cref{thm:balanced-two-row} is indexed by $t+1$ semistandard tableaux.  The Specht gauge basis may be chosen as standard tableaux of shape $(2t,t)$, and the Young--Jucys--Murphy footprint algebra measures that gauge label without resolving the logical semistandard-tableau label.

\begin{remark}[Rate and strength of the noise model]\label{rem:balanced-rate}
For the balanced three-colour sheet space and the full monodromy algebra, even the optimal logical dimension grows only linearly with $t$, and the rate $\log_3(t+1)/(3t)$ in the master qutrit count tends to zero.  In return, the ideal code corrects the full algebra generated by all permutations of $3t$ roots, not merely permutations of bounded length.  This is a protection statement for highly correlated monodromy noise, not a positive-rate code against generic local noise.
\end{remark}

\begin{remark}[Other partitions and colours]\label{rem:other-kostka}
Every pair $(\lambda,\boldsymbol m)$ with $K_{\lambda\boldsymbol m}>1$ gives a nontrivial cyclic code.  Varying the number of nodes, twist, degrees, and partition produces a large family whose logical dimensions are Kostka numbers.  The balanced three-cycle is singled out because its first member is a qubit and its multiplicity has the closed formula \eqref{eq:balanced-kostka}.  The balanced three-colour case is completely optimized above; formula \eqref{eq:optimal-dimension} gives the corresponding optimization criterion for other compositions.
\end{remark}

\section{Higher-rank quantizations, decoding, and implementation}\label{sec:higher-argyle}

In the argyle master space of Section~\ref{sec:argyle-master}, every node has rank one, and the exact error algebra comes from permutations of identical projective factors.  For argyle quivers with higher-rank nodes, bundle automorphism groups can be nonreductive, strata can have Grassmannian factors, and collisions can change the splitting type.  In this section we prove a quantization formula under explicit product-completion hypotheses, compute the code spaces for two group actions on the quantization of $\Gr(2,6)$, and state exact recovery criteria for specified collision operators.

\subsection{Quantization of a product completion}

Let $Q$ be an argyle block with fixed splitting type $\boldsymbol a$.  Motivated by the projective-space and Grassmannian descriptions in \cite[arXiv version, Theorems~4.1--4.2]{RayanSundboArgyle2018}, suppose that a chosen projective completion is supplied with a global product isomorphism
\begin{equation}\label{eq:argyle-product}
 Y_{Q,\boldsymbol a}
 \cong
 \Pp^q\times\prod_{\alpha=1}^N\Gr(s_\alpha,n_\alpha),
\end{equation}
for specified integers $q\geq0$ and $1\leq s_\alpha\leq n_\alpha$.  The global product identification is an explicit hypothesis in the calculation that follows.  Local quotient coordinates, or a projective bundle over a Grassmannian, do not by themselves supply it.  For decompositions into $(1,k_i,1)$ pieces, the stability conditions remain those induced from the full chain.

Choose a positive multi-polarization
\begin{equation}\label{eq:argyle-polarization}
 \mathscr A_{\boldsymbol\ell}
 =\cO_{\Pp^q}(\ell_0)
 \boxtimes
 \mathop{\boxtimes}_{\alpha=1}^N
 \cO_{\Gr(s_\alpha,n_\alpha)}(\ell_\alpha),
 \qquad \ell_\alpha>0.
\end{equation}

\begin{proposition}[Quantization of a polarized product completion]\label{prop:argyle-quantization}
With standard invariant Hermitian structures,
\begin{equation}\label{eq:argyle-quantization}
 H^0(Y_{Q,\boldsymbol a},\mathscr A_{\boldsymbol\ell})^*
 \cong
 \Sym^{\ell_0}(\C^{q+1})
 \otimes
 \bigotimes_{\alpha=1}^N
 \mathbb S_{(\ell_\alpha^{s_\alpha})}(\C^{n_\alpha}),
\end{equation}
using the convention that projective spaces parametrize lines and $\Gr(s,n)$ parametrizes subspaces.  The same tensor factorization holds over several blocks whenever both their chosen completion and their polarization split as external products.
\end{proposition}

\begin{proof}
K\"unneth's theorem factors the space of global sections over the product.  The projective-space factor gives the indicated symmetric power.  The Borel--Weil theorem \cite{FultonHarris1991} identifies the dual of the section space of the $\ell_\alpha$th Pl\"ucker power on $\Gr(s_\alpha,n_\alpha)$ with the irreducible rectangular Schur module.  The blockwise assertion is another application of K\"unneth under the stated product hypotheses.
\end{proof}

Equation \eqref{eq:argyle-quantization} gives explicit finite-dimensional quantizations, not merely dimensions.  Compact unitary groups act on their factors, and diagonal actions on repeated or coupled blocks admit multiplicity decompositions.  The same algebraic mechanism as \Cref{thm:cyclic-code} then applies.

\begin{corollary}[Multiplicity codes on argyle strata]\label{cor:argyle-multiplicity}
Suppose a compact group $G$ acts unitarily on the Hilbert space in \eqref{eq:argyle-quantization} and that
\begin{equation}\label{eq:argyle-isotypic}
 \cH_{Q,\boldsymbol a}
 \cong\bigoplus_\nu V_\nu\otimes M_\nu
\end{equation}
is its isotypic decomposition.  If the error algebra is the finite-dimensional
unital $*$-algebra generated by the unitary representation of $G$, then every
$M_\nu$ is a noiseless subsystem; fixing a vector in $V_\nu$ gives an exact
subspace code of dimension $\dim M_\nu$.
\end{corollary}

This corollary is representation-theoretic.  The polarized completion fixes the Hilbert space, while the chosen group action specifies the noise model; an action on coefficient coordinates is not automatically intrinsic to the quiver.  In the master-space case, $G=S_M$ and \eqref{eq:argyle-isotypic} is Schur--Weyl duality.  Higher-rank examples replace symmetric powers by Grassmannian Schur modules and their tensor-product multiplicities.

\subsection{A concrete higher-rank Grassmannian code}

An explicit two-node $A$-type stratum from \cite[arXiv version, Section~5]{RayanSundboArgyle2018} arises from the $\cO(6)$-twisted chain
\begin{equation}\label{eq:Gr26-quiver}
 \cO^{\oplus2}
 \xrightarrow{\ \Phi\ }
 \cO(-1)\otimes\cO(6)=\cO(5).
\end{equation}
This is a higher-rank endpoint chain in the same twisted-chain framework, rather than a three-node $(1,k,1)$ argyle block.  A stable map determines the two-plane spanned by its polynomial components in
\[
 V=H^0(\Pp^1,\cO(5))\cong\C^6,
\]
and quotienting by $\mathrm{GL}_2$ identifies the moduli stratum with $\Gr(2,V)$.  Here stability is equivalent to linear independence of the two polynomial components.  To see this, the total slope is $-1/3$.  Dependence gives an invariant constant line $\cO$ in the source kernel and hence instability.  Under independence, every subbundle $F$ of $E=\cO^{\oplus2}\oplus\cO(-1)$ has degree at most zero, since each summand in its Birkhoff--Grothendieck splitting has nonpositive degree.  If $\deg F=0$, then $F\cong\cO^{\oplus s}$ and its inclusion is a constant subspace of the source $\cO^{\oplus2}$, because $\Hom(\cO,\cO(-1))=0$.  Higgs invariance would force that constant subspace into the kernel of $\Phi$, contrary to linear independence.  Thus every proper invariant subbundle has degree at most $-1$ and rank at most two, giving slope at most $-1/2<-1/3$ and proving stability.  Common zeroes of the two forms do not affect this argument.  Its level-one Pl\"ucker quantization is
\begin{equation}\label{eq:Gr26-quantization}
 \cH_{\Gr}=H^0(\Gr(2,V),\cO(1))^*
 \cong\bigwedge^2V.
\end{equation}
The following theorem compares two actions of $\mathrm{SU}(2)$ on this quantization.

\Needspace{7\baselineskip}
\begin{theorem}[Two exact symmetry calculations on $\Gr(2,6)$]\label{thm:Gr26-codes}
For the quantization \eqref{eq:Gr26-quantization}, the following hold.
\begin{enumerate}
\item Let $A\cong\C^2$ be the homogeneous-coordinate representation of $\mathrm{SU}(2)$ on $\Pp^1$, so that $V\cong\Sym^5A$.  Under the induced base-rotation action,
\begin{equation}\label{eq:Gr26-natural-SU2}
 \bigwedge^2\Sym^5A
 \cong \Sym^8A\oplus\Sym^4A\oplus\C.
\end{equation}
The decomposition is multiplicity-free, so this natural $\mathrm{SU}(2)$ action has no protected multiplicity space of dimension greater than one.
\item Suppose instead that the six coefficient modes are supplied with a unitary tensor factorization $V=A\otimes B$, where $\dim A=2$ and $\dim B=3$, and let the error algebra be generated by $\mathrm{SU}(A)$ acting on the first factor.  Then
\begin{equation}\label{eq:Gr26-factorized-SU2}
 \bigwedge^2(A\otimes B)
 \cong
 \big(\textstyle\bigwedge^2A\otimes\Sym^2B\big)
 \oplus
 \big(\Sym^2A\otimes\textstyle\bigwedge^2B\big).
\end{equation}
The first summand is a six-dimensional decoherence-free subspace, and the second contains an exact logical qutrit with three-dimensional gauge factor $\Sym^2A$.  Both correct every channel whose Kraus operators lie in the algebra generated by this $\mathrm{SU}(A)$ action.
\end{enumerate}
\end{theorem}

\begin{proof}
The Clebsch--Gordan rule gives
\[
 \Sym^5A\otimes\Sym^5A
 \cong\bigoplus_{j=0}^{5}\Sym^{10-2j}A.
\]
The summand indexed by $j$ has exchange parity $(-1)^j$, so the exterior square contains exactly $j=1,3,5$, which proves \eqref{eq:Gr26-natural-SU2}.  Each irreducible occurs once.

For the factorized action, the degree-two Cauchy identity gives \eqref{eq:Gr26-factorized-SU2}.  The factor $\bigwedge^2A$ is the trivial one-dimensional representation of $\mathrm{SU}(A)$, while $\Sym^2A$ is its three-dimensional irreducible representation.  The corresponding multiplicity factors have dimensions
\[
 \dim\Sym^2B=6,
 \qquad
 \dim\bigwedge^2B=3.
\]
The exact correction statements follow from the same gauge-reset calculation as \Cref{thm:cyclic-code}.
\end{proof}

The second action uses a chosen grouping of the six polynomial modes into a $2\times3$ array; it is additional physical structure, not an invariant of the bare Grassmannian.  The contrast between \eqref{eq:Gr26-natural-SU2} and \eqref{eq:Gr26-factorized-SU2} is therefore a concrete demonstration that the moduli variety fixes the Hilbert space but does not, by itself, fix the error algebra.  Once the action is specified, the protected dimensions are exact.

\subsection{Collision strata and decoding criteria}

\paragraph{Invariant-zero collisions.}

Fix a $(1,k,1)$ block.  After Euclidean reduction by the nonreductive automorphism group, certain incoming and outgoing polynomial components retain gauge-invariant zeroes.  A type-changing collision occurs when the relevant collections share a zero.  Algebraically, the collision locus is cut out locally by resultants and, when several components are involved, by determinantal-resultantal ideals.

The following three levels should not be conflated.
\begin{enumerate}
\item The \emph{classical collision locus} is a subvariety of the quiver moduli space; it is a divisor only when it has codimension one.  Common-zero conditions involving several components can have higher codimension.
\item A \emph{collision observable} is a chosen Hermitian operator in a quantization of that moduli space.  A divisor does not canonically determine a projective measurement; a Hermitian metric, quantization level, and observable prescription are additional data.
\item A \emph{collision syndrome} is the outcome of such an observable only when its projectors are logically safe for the selected code and error family.
\end{enumerate}
For example, a resultant section $R_D$ defining a collision divisor, together with a Hermitian metric on its line bundle, gives a global squared-norm function $\|R_D\|^2$.  At a chosen quantization level, a normalized real function of this norm has a Hermitian Toeplitz quantization.  Its spectral projectors are diagnostic.  They become QEC syndrome projectors only after the compression conditions on the code have been checked.

The Birkhoff--Grothendieck splitting type is a coarser discrete footprint.  Let $\mathfrak S_Q$ be the finite set of stable splitting types reached by the type-change algorithm, and let $\cG_Q$ be the directed graph whose edges are the elementary moves \eqref{eq:type-change}.  A collision history is a path in $\cG_Q$ together with its continuous polynomial data; when reversing paths is discussed, the reverse edges are also included in the history model.

\paragraph{A stratified Knill--Laflamme criterion.}

To model errors between splitting types, choose a finite-dimensional Hilbert space $\cH_{\boldsymbol a}$ for every $\boldsymbol a\in\mathfrak S_Q$ and set
\begin{equation}\label{eq:stratified-Hilbert}
 \cH_{\mathrm{strat}}
 =\bigoplus_{\boldsymbol a\in\mathfrak S_Q}\cH_{\boldsymbol a}.
\end{equation}
Let $\Pi_{\boldsymbol a}$ be the summand projectors.  These choices are part of the model: distinct algebraic strata are not automatically orthogonal quantum sectors.

Fix an initial type $\boldsymbol a_0$, a code $\cC\subset\cH_{\boldsymbol a_0}$ with projector $P$, and a family of collision-history operators $E_p$ indexed by paths $p$ beginning at $\boldsymbol a_0$.  Assume that $E_p\cC$ lies in the summand associated with the terminal type $t(p)$.

\begin{theorem}[Argyle collision-fibre criterion]\label{thm:argyle-fibre-KL}
Under the preceding assumptions, measurement of the type projectors $\{\Pi_{\boldsymbol a}\}$ followed by a type-conditioned recovery exactly corrects the collision-history family $\{E_p\}$ if and only if, for every terminal type $\boldsymbol b$,
\begin{equation}\label{eq:argyle-fibre-KL}
 PE_p^\dagger E_qP=\lambda_{pq}^{(\boldsymbol b)}P
 \qquad
 \text{whenever }t(p)=t(q)=\boldsymbol b.
\end{equation}
Histories with different terminal types have vanishing cross-compressions automatically.
\end{theorem}

\begin{proof}
If $t(p)\neq t(q)$, the ranges $E_p\cC$ and $E_q\cC$ lie in orthogonal summands of \eqref{eq:stratified-Hilbert}; hence $PE_p^\dagger E_qP=0$.  Within a fixed terminal summand, the ordinary Knill--Laflamme theorem gives exactly \eqref{eq:argyle-fibre-KL}.  Measuring the summand first makes the cross-summand alternative classical, after which the standard recovery is applied independently in each fibre.  Conversely, any exact type-conditioned recovery implies the within-type Knill--Laflamme equations.
\end{proof}

The theorem turns the type-change graph into a footprint graph.  The endpoint is measured; the path remains latent.  Two collision histories with the same endpoint may differ by a closed path or by continuous data internal to the stratum.  Their residual composite is harmless exactly when \eqref{eq:argyle-fibre-KL} is scalar.  The theorem is conditional on the specified Hilbert sectors and collision operators; it does not construct them from the type-change graph.  Its algebraic mechanism is illustrated by the cyclic codes when errors send the prepared gauge vector into definite tableau sectors.  General monodromies may create superpositions of those sectors, although gauge reset still corrects them.

\paragraph{Geometric likelihoods on the type-change graph.}

The type-change graph also organizes representative-based decoding.  Assign an edge $e$ a cost $w(e)$ derived from a physical noise model, a norm of the polynomial perturbation, a vanishing order of a resultant, or a local action.  Given a measured terminal type $\boldsymbol b$ and an admissible set of histories on which the minimum is attained, a minimum-cost decoder chooses
\begin{equation}\label{eq:argyle-decoder}
 \widehat p(\boldsymbol b)
 \in\operatorname*{argmin}_{p:\,t(p)=\boldsymbol b}
 \left(\sum_{e\in p}w(e)+w_{\mathrm{cont}}(p)\right).
\end{equation}
This is a moduli-dependent analogue of choosing a minimum-weight error chain with fixed boundary.  The following calculation isolates exactly what such a choice accomplishes.

\begin{proposition}[Residual of a representative decoder]\label{prop:representative-decoder}
Assume the collision-history operators $E_p$ are unitary on $\cH_{\mathrm{strat}}$, with $E_p\cC\subseteq\cH_{t(p)}$ as above.  After measuring $\boldsymbol b$, let the representative decoder apply $E_{\widehat p(\boldsymbol b)}^\dagger$.  If the actual path was $p$ with $t(p)=\boldsymbol b$, the compressed residual on the code is
\begin{equation}\label{eq:representative-residual}
 L_{p\mid\boldsymbol b}
 =P E_{\widehat p(\boldsymbol b)}^\dagger E_pP.
\end{equation}
The decoder restores every encoded state after the history $p$ if and only if $L_{p\mid\boldsymbol b}$ is a scalar phase times $P$.  It restores the entire endpoint fibre if and only if all the residuals \eqref{eq:representative-residual} are scalar phases; this condition is independent of the chosen representative.  It is stronger than the general criterion \eqref{eq:argyle-fibre-KL}, because a Knill--Laflamme recovery may further distinguish orthogonal error subspaces inside one terminal-type sector.
\end{proposition}

\begin{proof}
The representative inverse changes the actual error $E_p$ into
$E_{\widehat p(\boldsymbol b)}^\dagger E_p$.  Its restriction to the code is logically trivial exactly when its compression is a scalar of modulus one.  Unitarity then forces equality with that phase on the code, since the projected component already has full norm.  If this holds relative to one representative, then changing representatives only multiplies the residuals by phases.  By contrast, \eqref{eq:argyle-fibre-KL} permits scalar overlaps of modulus strictly less than one, including zero, and the corresponding recovery need not be a single representative inverse.
\end{proof}

Costs or posterior probabilities decide which representative is most plausible; \Cref{prop:representative-decoder} decides whether choosing incorrectly is logically harmless.  Comparable paths are therefore benign when their residual is scalar and dangerous when it lies nontrivially in the logical commutant.  Near an intersection of collision strata, instability of the measured endpoint is a distinct classical classification error and should not be conflated with the within-fibre logical residual.

\subsection{Encoding, measurement, and recovery}\label{sec:implementation}

\paragraph{An ideal protocol.}

For a cyclic composition $\boldsymbol m$ and a partition $\lambda$ with $K_{\lambda\boldsymbol m}>1$, the exact protocol has the following form.
\begin{enumerate}
\item \emph{Choose the regular fibre.}  Fix $\gamma\in\cB_M^{\reg}$ and a reference ordering of its roots.  The ordering is a coordinate choice for the sheet register; changing it conjugates the representation.
\item \emph{Prepare the sheet space.}  Use either the intrinsic basis $\ket{B_1,\ldots,B_r}$ or the word basis of the argyle weight sector $\cK_{M,r}[\boldsymbol m]$.
\item \emph{Transform to representation registers.}  Apply the Schur transform, restricted to the fixed weight sector, to obtain registers $(\lambda,T,S)$: an isotypic label $\lambda$, a standard-tableau gauge label $T$, and a logical weight-basis label $S$.  A compatible orthonormal Gelfand--Tsetlin basis can be indexed by semistandard tableaux of content $\boldsymbol m$; an arbitrary algebraic tableau basis need not be orthonormal.
\item \emph{Encode.}  Fix $\lambda$ and a gauge tableau $T_0$, and encode the state in the $K_{\lambda\boldsymbol m}$-dimensional $S$ register.
\item \emph{Allow monodromy noise.}  A root-braid history acts through $S_M$ on the gauge register and as the identity on $S$.
\item \emph{Measure an optional footprint.}  Young--Jucys--Murphy observables or compatible collision-parity projectors probe the gauge register without measuring $S$.
\item \emph{Recover.}  Discard the disturbed gauge register and reprepare $T_0$.  No estimate of the full root permutation is required.
\end{enumerate}

Efficient quantum circuits for the Schur transform are known with complexity polynomial in $M$, $r$, and $\log(1/\varepsilon)$ for approximation accuracy $\varepsilon$ \cite{BaconChuangHarrow2006}.  This means that the representation-theoretic change of basis is not, by itself, an exponential obstruction.  Preparing a particular quiver-moduli sheet model and realizing its monodromy channel are separate physical questions.

\paragraph{Footprint-assisted recovery versus passive protection.}

The codes are noiseless subsystems for the ideal monodromy algebra.  Consequently, footprint measurement is optional: tracing out and resetting the gauge factor recovers the logical state without learning which monodromy occurred.  A measured footprint becomes useful in three situations.
\begin{enumerate}
\item The implementation uses outcome-dependent gauge preparation in place of a direct reset channel.  A projective footprint measurement is not, in general, a coherent restoration of the pre-error gauge state.
\item The error model is only approximately contained in the monodromy algebra, and the measured sector selects a calibrated recovery.
\item The measurement record is used for likelihood estimation, diagnostics, or architecture control even though the logical subsystem is passively protected.
\end{enumerate}
This distinction mirrors operator QEC more generally: a syndrome may guide active recovery, but a noiseless subsystem can make the logical information independent of the syndrome.

\paragraph{Relation to permutation-invariant codes.}

Permutation-invariant codes encode in a symmetric subspace.  Ouyang constructs families that correct bounded-weight errors exactly and others that correct spontaneous-decay errors approximately \cite{Ouyang2014}.  Ouyang and Brennen give a recovery algorithm for deletions in certain permutation-invariant codes \cite{OuyangBrennen2026}.  Here permutations themselves are the geometrically selected errors.  The logical information is stored in multiplicity spaces on which those permutations act trivially.  The two uses of symmetric-group representation theory are therefore almost dual: one uses symmetry of the states, while the other uses the commutant of the error action.

\paragraph{Error-model boundary and a perturbative test.}

Exact correction depends only on the restriction of the errors to the selected isotypic block.  Thus global membership in the group-algebra image is sufficient but not necessary.

\begin{proposition}[Blockwise extension of the exact error model]\label{prop:blockwise-extension}
Fix a block $S^\lambda\otimes\cL_{\lambda,\boldsymbol m}$ and a gauge vector $v$.  Let $\{E_a\}$ be any error family that preserves this block and satisfies
\begin{equation}\label{eq:blockwise-gauge-errors}
 E_a\big|_\lambda=B_a\otimes I_{\cL}
\end{equation}
for arbitrary $B_a\in\End(S^\lambda)$.  Then the gauge-fixed code $\C v\otimes\cL_{\lambda,\boldsymbol m}$ exactly corrects $\{E_a\}$, whether or not the operators have been specified on the other isotypic blocks.
\end{proposition}

\begin{proof}
For the code projector $P_v=\proj v\otimes I_{\cL}$,
\[
 P_vE_a^\dagger E_bP_v
 =\bra vB_a^\dagger B_b\ket v\,P_v.
\]
These are the Knill--Laflamme equations.
\end{proof}

For components that do not have the form \eqref{eq:blockwise-gauge-errors}, the departure from exactness can be bounded directly.  Let $P$ denote the gauge-fixed code projector and $P_\lambda$ the central isotypic projector \eqref{eq:central-isotypic-projector}.

\begin{proposition}[Perturbative Knill--Laflamme defect]\label{prop:perturbative-KL}
Write each error operator as
\begin{equation}\label{eq:error-perturbation}
 E_a=A_a+\Delta_a,
 \qquad A_a\in\cA_{\mathrm{mon}}(\boldsymbol m),
 \qquad \varepsilon_a=\|\Delta_a\|.
\end{equation}
Choose $\lambda_{ab}$ by $PA_a^\dagger A_bP=\lambda_{ab}P$.  Then
\begin{equation}\label{eq:KL-defect-bound}
 \big\|PE_a^\dagger E_bP-\lambda_{ab}P\big\|
 \leq
 \|A_a\|\varepsilon_b
 +\varepsilon_a\|A_b\|
 +\varepsilon_a\varepsilon_b.
\end{equation}
Moreover, leakage from the selected isotypic block obeys
\begin{equation}\label{eq:isotypic-leakage-bound}
 \|(I-P_\lambda)E_aP\|\leq\varepsilon_a.
\end{equation}
If the $A_a$ are contractions, the right-hand side of \eqref{eq:KL-defect-bound} is at most $\varepsilon_a+\varepsilon_b+\varepsilon_a\varepsilon_b$.
\end{proposition}

\begin{proof}
Expanding $E_a^\dagger E_b$ and subtracting the exact monodromy compression leaves
\[
 P\big(A_a^\dagger\Delta_b+\Delta_a^\dagger A_b
       +\Delta_a^\dagger\Delta_b\big)P.
\]
The operator-norm triangle inequality and submultiplicativity give \eqref{eq:KL-defect-bound}.  Since $A_a$ preserves every isotypic block and $P\leq P_\lambda$,
\[
 (I-P_\lambda)E_aP=(I-P_\lambda)\Delta_aP,
\]
which proves \eqref{eq:isotypic-leakage-bound}.
\end{proof}

This estimate cleanly separates the exact geometric component from unmodelled noise.  Berry phases from pure braids, nonadiabatic transitions near the discriminant, local errors in the master qutrits, and type-changing leakage all contribute to $\Delta_a$ unless their action is explicitly incorporated into a larger blockwise gauge algebra.  Equation \eqref{eq:KL-defect-bound} is therefore a direct calibration target: it measures how far the realized operator products depart from the scalar compressions required by the code.

\begin{corollary}[Operational perturbation bound]\label{cor:operational-perturbation}
Let $\mathcal N(\rho)=\sum_a E_a\rho E_a^\dagger$ and $\mathcal M(\rho)=\sum_a A_a\rho A_a^\dagger$ be trace-preserving channels with a common finite Kraus index set and $A_a\in\cA_{\mathrm{mon}}$.  Let $V$ encode a gauge-fixed code, put $P=VV^\dagger$, and write $\mathcal V(\rho)=V\rho V^\dagger$.  For the reset recovery $\mathcal R$ in \eqref{eq:reset-recovery}, with its trace-preserving extension,
\begin{equation}\label{eq:diamond-recovery-bound}
 \|\mathcal R\circ\mathcal N\circ\mathcal V-\mathcal V\|_\diamond
 \leq\min\{2,2\delta\},\qquad
 \delta=\left(\sum_a\|(E_a-A_a)P\|^2\right)^{1/2}.
\end{equation}
Here the diamond norm tests the channel on inputs entangled with an arbitrary reference system.  In particular, the trace distance, with its factor $1/2$, between the recovered and ideal encoded outputs is at most $\min\{1,\delta\}$.
\end{corollary}

\begin{proof}
The paired Stinespring isometries on the logical input are
\[
 T_E=\sum_a E_aV\otimes\ket a,
 \qquad T_A=\sum_a A_aV\otimes\ket a.
\]
Orthogonality of the environment basis gives $\|T_E-T_A\|\leq\delta$.  This norm bound is unchanged upon adjoining a reference system.  Expanding $T_E\rho T_E^\dagger-T_A\rho T_A^\dagger$ in two terms, and using that both maps are isometries, bounds the induced trace-norm difference by $2\delta$.  Partial trace and the recovery are channels and contract trace norm; moreover $\mathcal R\circ\mathcal M\circ\mathcal V=\mathcal V$.  These observations prove the $2\delta$ bound, while the distance between two channels is at most two.
\end{proof}

The additional normalization hypothesis in this corollary is essential: the algebra components $A_a$ in \Cref{prop:perturbative-KL} need not themselves form a channel.  A small pairwise Knill--Laflamme defect alone is not the operational bound \eqref{eq:diamond-recovery-bound}.  The latter uses a specified, normalized ideal channel and a pairing of its Kraus operators with those of the realized channel.

\subsection{The field-theoretic footprint dictionary}\label{sec:dictionary}

The diagrammatic field-theory framework of \cite{RayanDiagrammatic2026} separates the spatial or moduli datum, the code space, the error history, the measured footprint, the footprint fibre, and the residual logical class.  In the cyclic--argyle model these entries are summarized in \Cref{tab:dictionary}.

The discriminant is a locus in the Hitchin base.  A loop around it gives an error history; the measured footprint can be a transposition parity or a tableau label.  To use such a measurement as a syndrome, its projectors must be syndrome-admissible for the chosen code and error representatives.

The same distinction applies to gauge equivalence.  Two polynomial matrices related by bundle automorphisms represent the same moduli point and should not be decoded as different physical histories unless the realization retains the frame.  The root-allocation construction performs this quotient before forming the Hilbert space.  This is analogous to identifying stabilizer-equivalent Pauli representatives before asking whether the remaining difference is logical.

\begin{table}[H]
\centering
\small
\renewcommand{\arraystretch}{1}
\begin{tabular}{@{}>{\raggedright\arraybackslash}p{0.26\textwidth}@{\hspace{1.5em}}>{\raggedright\arraybackslash}p{0.67\textwidth}@{}}
\textbf{QEC data} & \textbf{Quiver model}\\[3pt]
\hline
\noalign{\vskip 5pt}
Geometry & Regular cyclic Hitchin-base point $\gamma$ and argyle master space $X_{M,r}$\\[4pt]
State space & Sheet space $\ell^2(h_Q^{-1}(\gamma))$, or weight sector $\cK_{M,r}[\boldsymbol m]$\\[4pt]
Logical space & Kostka multiplicity space $\mathbb S_\lambda(\C^r)[\boldsymbol m]$\\[4pt]
Gauge space & Specht module $S^\lambda$\\[4pt]
Error history & Root braid and its permutation monodromy; specified operator for an argyle type-change path\\[4pt]
Footprint & Collision parity, tableau content spectrum, resultant data, or terminal splitting type\\[4pt]
Syndrome algebra & Logically safe commuting gauge algebra, syndrome-admissible for sector-resolved representatives\\[4pt]
Footprint fibre & Monodromies or collision paths with the same measured gauge or type data\\[4pt]
Harmless residual & Monodromy-algebra operator acting only on the Specht gauge factor\\[4pt]
Logical residual & Non-scalar commutant action on the Kostka factor\\[4pt]
Recovery & Trace out and reset gauge, or measure type and apply fibrewise Knill--Laflamme recovery\\
\end{tabular}
\caption{QEC data in the cyclic and argyle models.}
\label{tab:dictionary}
\end{table}

\section{Overall Summary}

The node-labelled cyclic Hitchin fibres yield exact codes for root-permutation noise.  Specht factors carry the noise, Kostka factors carry the logical state, and gauge reset supplies recovery.  The largest Kostka multiplicity bounds every exact subspace or subsystem encoding for the full monodromy algebra, even across isotypic blocks.

The argyle master space realizes these sheet modules as tensor-power weight sectors, with an equivariant bridge descending through the root-ordering cover.  For the balanced three-cycle, the complete spectrum \eqref{eq:balanced-full-kostka} proves the optimal logical dimension $t+1$.  The six-sheet qubit's right group algebra realizes every logical unitary, producing same-footprint errors that no common recovery can correct.

The $\Gr(2,6)$ calculation exhibits the dependence on the chosen noise action.  The stratified criterion \eqref{eq:argyle-fibre-KL} characterizes recovery for specified collision operators and type sectors.  Finally, \eqref{eq:KL-defect-bound} bounds algebraic departures from the ideal model, and \eqref{eq:diamond-recovery-bound} bounds recovery error relative to a normalized ideal channel.

\section*{Acknowledgements}

The first-named author acknowledges partial support from a Natural Sciences and Engineering Research Council of Canada (NSERC) Discovery Grant.  He thanks Syddansk Universitet (SDU), the Centre for Quantum Mathematics at SDU, the Danish e-infrastructure Consortium (DeiC), and the Danish Institute for Advanced Study for hosting him in August 2026 during the Niels Bohr Quantum Summer School and for providing an inspiring and comfortable environment in which to complete this manuscript.

\begingroup
\small
\raggedright

\endgroup

\end{document}